\documentclass[journal]{IEEEtran}
\usepackage{url}
\usepackage[utf8]{inputenc}
\usepackage{xcolor}
\usepackage{amsmath}
\usepackage{amssymb}
\usepackage{amsmath}
\DeclareMathOperator*{\argmax}{arg\,max}

\usepackage[acronyms,nonumberlist,nopostdot,nomain,nogroupskip]{glossaries}
\usepackage{tablefootnote}
\usepackage{booktabs}
\usepackage{tabularx}
\usepackage{tikz}
\usepackage{pgfplots}
\pgfplotsset{compat=newest} 
\pgfplotsset{plot coordinates/math parser=false} 
\newlength\fheight
\newlength\fwidth
\usetikzlibrary{plotmarks,patterns,decorations.pathreplacing,backgrounds,calc,arrows,arrows.meta,spy,matrix}
\usepgfplotslibrary{patchplots,groupplots}
\usepackage{tikzscale}
\usepackage{siunitx}

\usepackage{multirow}
\usepackage{tkz-kiviat}

\usepackage[font=footnotesize]{subcaption}
\usepackage[font=small]{caption}

\usepackage{mathtools}

\usepackage{dblfloatfix}    
\usepackage{colortbl}

\usepackage{makecell}
\usepackage{diagbox}
\usepackage{tikz-qtree}
\usetikzlibrary{trees} 

\newacronym{3gpp}{3GPP}{3rd Generation Partnership Project}
\newacronym{adc}{ADC}{Analog to Digital Converter}
\newacronym{5g}{5G}{5th generation}
\newacronym{6g}{6G}{6th generation}
\newacronym{aimd}{AIMD}{Additive Increase Multiplicative Decrease}
\newacronym{am}{AM}{Acknowledged Mode}
\newacronym{amc}{AMC}{Adaptive Modulation and Coding}
\newacronym{aqm}{AQM}{Active Queue Management}
\newacronym{awgn}{AGWN}{Additive White Gaussian Noise}
\newacronym{balia}{BALIA}{Balanced Link Adaptation}
\newacronym{bdp}{BDP}{Bandwidth-Delay Product}
\newacronym{bf}{BF}{beamforming}
\newacronym{cc}{CC}{Congestion Control}
\newacronym{cdf}{CDF}{Cumulative Distribution Function}
\newacronym{pdf}{PDF}{Probability Density Function}
\newacronym{cn}{CN}{Core Network}
\newacronym{cqi}{CQI}{Channel Quality Information}
\newacronym{cp}{CP}{Control Plane}
\newacronym{csirs}{CSI-RS}{Channel State Information - Reference Signal}
\newacronym{dc}{DC}{Dual Connectivity}
\newacronym{rb}{RB}{Resource Block}
\newacronym{dce}{DCE}{Direct Code Execution}
\newacronym{dci}{DCI}{Downlink Control Information}
\newacronym{udp}{UDP}{User Datagram Protocol}
\newacronym{dl}{DL}{Downlink}
\newacronym{dmr}{DMR}{Deadline Miss Ratio}
\newacronym{dmrs}{DMRS}{DeModulation Reference Signal}
\newacronym{e2e}{E2E}{End-to-End}
\newacronym{ppp}{PPP}{Poission Point Process}
\newacronym{si}{SI}{Study Item}
\newacronym{ecn}{ECN}{Explicit Congestion Notification}
\newacronym{edf}{EDF}{Earliest Deadline First}
\newacronym{enb}{eNB}{eNodeB}
\newacronym{epc}{EPC}{Evolved Packet Core}
\newacronym{es}{ES}{Edge Server}
\newacronym{cav}{CAV}{Connected and Autonomous Vehicle}
\newacronym{fdma}{FDMA}{Frequency Division Multiple Access}
\newacronym{fdd}{FDD}{Frequency Division Duplexing}
\newacronym{upa}{UPA}{Uniform Planar Array}
\newacronym[firstplural=Radio Access Technologies (RATs)]{rat}{RAT}{Radio Access Technology}
\newacronym[firstplural=Radio Access Technology (RTs)]{rt}{RT}{Radio Technology}
\newacronym{fs}{FS}{Fast Switching}
\newacronym{isd}{ISD}{inter-site distance}
\newacronym{ftp}{FTP}{File Transfer Protocol}
\newacronym{gnb}{gNB}{Next Generation Node Base}
\newacronym{harq}{HARQ}{Hybrid Automatic Repeat reQuest}
\newacronym{hetnet}{HetNet}{Heterogeneous Network}
\newacronym{hh}{HH}{Hard Handover}
\newacronym{hol}{HOL}{Head-of-Line}
\newacronym{ia}{IA}{Initial Access}
\newacronym{imt}{IMT}{International Mobile Telecommunication}
\newacronym{iot}{IoT}{Internet of Things}
\newacronym{los}{LOS}{Line of Sight}
\newacronym{lte}{LTE}{Long Term Evolution}
\newacronym{m2m}{M2M}{Machine to Machine}
\newacronym{mac}{MAC}{Medium Access Control}
\newacronym{mc}{MC}{Multi-Connectivity}
\newacronym{mcs}{MCS}{Modulation and Coding Scheme}
\newacronym{mec}{MEC}{Mobile Edge Cloud}
\newacronym{mi}{MI}{Mutual Information}
\newacronym{mimo}{MIMO}{Multiple Input Multiple Output}
\newacronym{mmwave}{mmWave}{millimeter wave}
\newacronym{mptcp}{MPTCP}{Multipath TCP}
\newacronym{mr}{MR}{Maximum Rate}
\newacronym{mss}{MSS}{Maximum Segment Size}
\newacronym{mtd}{MTD}{Machine-Type Device}
\newacronym{mtu}{MTU}{Maximum Transmission Unit}
\newacronym{nfv}{NFV}{Network Function Virtualization}
\newacronym{vnf}{VNF}{ Virtualization Network Function}
\newacronym{sdn}{SDN}{Software Defined Networking}
\newacronym{nlos}{NLOS}{Non Line of Sight}
\newacronym{nlosb}{NLOSb}{Building Non Line of Sight}
\newacronym{nlosv}{NLOSv}{Vehicle Non Line of Sight}
\newacronym{nr}{NR}{New Radio}
\newacronym{ofdm}{OFDM}{Orthogonal Frequency Division Multiplexing}
\newacronym{pdcch}{PDCCH}{Physical Downlonk Control Channel}
\newacronym{pdcp}{PDCP}{Packet Data Convergence Protocol}
\newacronym{pdsch}{PDSCH}{Physical Downlink Shared Channel}
\newacronym{pdu}{PDU}{Packet Data Unit}
\newacronym{pf}{PF}{Proportional Fair}
\newacronym{pgw}{PGW}{Packet Gateway}
\newacronym{phy}{PHY}{Physical}
\newacronym{pbch}{PBCH}{Physical Broadcast Channel}
\newacronym[plural=\gls{mme}s,firstplural=Mobility Management Entities (MMEs)]{mme}{MME}{Mobility Management Entity}
\newacronym{prb}{PRB}{Physical Resource Block}
\newacronym{pss}{PSS}{Primary Synchronization Signal}
\newacronym{pucch}{PUCCH}{Physical Uplink Control Channel}
\newacronym{pusch}{PUSCH}{Physical Uplink Shared Channel}
\newacronym{rach}{RACH}{Random Access Channel}
\newacronym{ran}{RAN}{Radio Access Network}
\newacronym{red}{RED}{Random Early Detection}
\newacronym{rf}{RF}{Radio Frequency}
\newacronym{rlc}{RLC}{Radio Link Control}
\newacronym{rlf}{RLF}{Radio Link Failure}
\newacronym{rrc}{RRC}{Radio Resource Control}
\newacronym{rrm}{RRM}{Radio Resource Management}
\newacronym{rr}{RR}{Round Robin}
\newacronym{rs}{RS}{Remote Server}
\newacronym{rsrp}{RSRP}{Reference Signal Received Power}
\newacronym{rss}{RSS}{Received Signal Strength}
\newacronym{rtt}{RTT}{Round Trip Time}
\newacronym{rw}{RW}{Receive Window}
\newacronym{rx}{RX}{Receiver}
\newacronym{sa}{SA}{standalone}
\newacronym{sack}{SACK}{Selective Acknowledgment}
\newacronym{sap}{SAP}{Service Access Point}
\newacronym{sch}{SCH}{Secondary Cell Handover}
\newacronym{scoot}{SCOOT}{Split Cycle Offset Optimization Technique}
\newacronym{sdma}{SDMA}{Spatial Division Multiple Access}
\newacronym{sinr}{SINR}{Signal to Interference plus Noise Ratio}
\newacronym{sm}{SM}{Saturation Mode}
\newacronym{snr}{SNR}{Signal to Noise Ratio}
\newacronym{son}{SON}{Self-Organizing Network}
\newacronym{ss}{SS}{Synchronization Signal}
\newacronym{srs}{SRS}{Sounding Reference Signal}
\newacronym{sss}{SSS}{Secondary Synchronization Signal}
\newacronym{tb}{TB}{Transport Block}
\newacronym{tcp}{TCP}{Transmission Control Protocol}
\newacronym{tdd}{TDD}{Time Division Duplexing}
\newacronym{tdma}{TDMA}{Time Division Multiple Access}
\newacronym{tfl}{TfL}{Transport for London}
\newacronym{tm}{TM}{Transparent Mode}
\newacronym{prr}{PRR}{Packet Reception Ratio}
\newacronym{trp}{TRP}{Transmitter Receiver Pair}
\newacronym{tti}{TTI}{Transmission Time Interval}
\newacronym{ttt}{TTT}{Time-to-Trigger}
\newacronym{tx}{TX}{Transmitter}
\newacronym{ue}{UE}{User Equipment}
\newacronym{ul}{UL}{Uplink}
\newacronym{uml}{UML}{Unified Modeling Language}
\newacronym{um}{UM}{Unacknowledged Mode}
\newacronym{utc}{UTC}{Urban Traffic Control}
\newacronym{vm}{VM}{Virtual Machine}
\newacronym{rsrq}{RSRQ}{Reference Signal Received Quality}
\newacronym{rssi}{RSSI}{Received Signal Strength Indicator}
\newacronym{crs}{CRS}{Cell Reference Signal}
\newacronym{v2v}{V2V}{Vehicle-to-Vehicle}
\newacronym{v2i}{V2I}{Vehicle-to-Infrastructure}
\newacronym{v2n}{V2N}{Vehicle-to-Network}
\newacronym{v2x}{V2X}{Vehicle-to-Everything}
\newacronym{vn}{VN}{Vehicular Node}
\newacronym{dsrc}{DSRC}{Dedicated Short Range Communication}
\newacronym{ci}{CI}{context information}
\newacronym{voi}{VoI}{value of information}
\newacronym{gps}{GPS}{Global Positioning System}
\newacronym{qos}{QoS}{Quality of Service}
\newacronym{qoe}{QoE}{Quality of Experience}
\newacronym{ml}{ML}{Machine Learning}
\newacronym{ahp}{AHP}{Analytic Hierarchy Process}
\newacronym{lidar}{LIDAR}{Light Detection and Ranging}
\newacronym{sumo}{SUMO}{Simulation of Urban MObility}
\newacronym{wave}{WAVE}{Wireless Access in Vehicular Environment}
\newacronym{c-its}{C-ITS}{Connected Intelligent Transportation System}
\newacronym{dash}{DASH}{Dynamic Adaptive Streaming over HTTP}
\newacronym{http}{HTTP}{HyperText Transfer Protocol}
\newacronym{nt}{NT}{non-terrestrial}
\newacronym{ntc}{NTC}{non-terrestrial communication}
\newacronym{haps}{HAPS}{High Altitude Platform Station}
\newacronym{leo}{LEO}{Low Earth Orbit}
\newacronym{meo}{MEO}{Medium Earth Orbit}
\newacronym{geo}{GEO}{Geostationary Earth Orbit}
\newacronym{uav}{UAV}{Unmanned Aerial Vehicle}
\newacronym{nsat}{nSAT}{Nanosatellite}
\newacronym{ehf}{EHF}{extremely high-frequency}
\newacronym{ioe}{IoE}{Internet of Everyone}
\newacronym{gan}{GaN}{Gallium Nitride}
\newacronym{aoi}{AoI}{Area of Interest}
\usepackage{cite}
\usepackage{bbold}
\usepackage{bbm}

\usepackage{array}
\newcolumntype{?}{!{\vrule width 1.5pt}}
\newcolumntype{P}[1]{>{\centering\arraybackslash}p{#1}}
\usepackage{makecell}
\usepackage{mdframed}
\usepackage[many]{tcolorbox}
\usepackage{enumitem}

\usepackage{ntheorem}
\theoremheaderfont{\hspace*{\parindent}\itshape }
\theorembodyfont{\normalfont}
\theoremseparator{:}

\theoremheaderfont{\hspace*{\parindent}\bfseries}
\theorembodyfont{\normalfont}
\theoremseparator{:}
\theoremstyle{remark}

 \def \squeeze_f_i{
    \begin{equation}\label{fi_lemma_value}
        \begin{aligned}
        f_i(r) &= 2\pi\lambda_{\rm UAV}rp_i(b(r))\exp\left(-2\pi\lambda_{\rm UAV}\int_0^{b(r)}p_i(\rho)\rho d\rho\right)
    \end{aligned}
    \end{equation}
            }

\theoremheaderfont{\hspace*{\parindent} \normalfont \bfseries}
\theorembodyfont{\itshape}
\theoremseparator{:}
\newtheorem{theorem}{Theorem}
\newtheorem{lemma}{Lemma}

\newtcbox{\mybox}[1][]{nobeforeafter,math upper,tcbox raise base,
  enhanced,frame hidden,boxrule=0pt,interior style={top color=green!10!white,
  bottom color=green!10!white,middle color=green!50!yellow},
  fuzzy halo=1pt with green,drop large lifted shadow,#1}

\usepackage{siunitx}
\usetikzlibrary{fadings}

\tikzfading[name=middle,
            top color=transparent!100,
            bottom color=transparent!100,
            middle color=transparent!20]

\usetikzlibrary{arrows,automata,calc,shapes, positioning,shadows,shadows.blur,shapes.geometric}

\makeglossaries

\begin{document}
\pagenumbering{gobble}


\title{Coverage Analysis of UAVs in Millimeter Wave Networks: A Stochastic Geometry Approach}
\author{\IEEEauthorblockN{ Matilde Boschiero, Marco Giordani, Michele Polese, Michele Zorzi}
\IEEEauthorblockA{\\
Department of Information Engineering (DEI), University of Padova, Italy\\
Email: {\texttt{\{name.surname\}@dei.unipd.it}}
\vspace{-0.6cm}}

\thanks{This work was partially supported by NIST through Award No. 70NANB17H166.}
}

\maketitle

\begin{abstract}
Recent developments in robotics and communication technologies are paving the way towards the use of Unmanned Aerial Vehicles (UAVs) to provide ubiquitous connectivity in public safety scenarios or in remote areas. The millimeter wave (mmWave) spectrum, in particular, has gained momentum since the huge amount of free spectrum available at such frequencies can yield very high data rates. In the UAV context, however, mmWave operations may incur severe signal attenuation and sensitivity to blockage, especially considering the very long transmission distances involved. In this paper, we present a tractable stochastic analysis to characterize the coverage probability of UAV stations operating at mmWaves. We exemplify some of the trade-offs to be considered when designing solutions for millimeter wave (mmWave) scenarios, such as the beamforming configuration, and the UAV altitude and deployment. 
\end{abstract}

\begin{picture}(0,0)(0,-300)
\put(0,0){
\put(0,0){This paper has been submitted to IWCMC 2020. Copyright may be transferred without notice.}}
\end{picture}

\begin{IEEEkeywords}
5G;  Unmanned Aerial Vehicles (UAVs); millimeter waves (mmWaves); stochastic geometry; coverage analysis.
\end{IEEEkeywords}

\section{Introduction} 
\label{sec:intro}
In recent years the usage of \glspl{uav}, commonly known as drones,  has rapidly grown thanks to the extremely low operating and maintenance costs, and to the ease of deployment. 
When  equipped with dedicated sensors, \glspl{uav} can support several services, from airspace surveillance and border patrol~\cite{chandhar2017massive} to traffic and crowd monitoring~\cite{chandhar2019massive}. 
Recently, drones have been studied as a solution to provide coverage and connectivity to ground users and first responders in emergency situations~\cite{shi2018drone}, e.g., when  cellular infrastructures are either unavailable or no longer operational \cite{mezzavilla2018public}. 
UAVs can also be deployed on-demand, to boost base station's capacity  in hot-spot areas or when terrestrial infrastructures are overloaded~\cite{giordani2020satellite}.

Today, UAV communications are typically enabled by legacy wireless technologies such as \gls{lte}~\cite{van2016lte} or Wi-Fi which, however, cannot satisfy the  very strict reliability, throughput and latency requirements of future  applications~\cite{hayat2015experimental}.
In these regards, \gls{5g} innovations, especially network operations in the \gls{mmwave} spectrum,  may offer a practical solution to overcome existing cellular connectivity shortfalls~\cite{zhang2019survey}.
In fact, the large spectrum available at \glspl{mmwave}, in combination with massive \gls{mimo} technologies,  makes it possible to achieve multi-Gbps transmission speeds, as well as to guarantee spatial isolation and immunity to jamming and eavesdropping through directional communication~\cite{xiao2017millimeter}.

Nevertheless, the application of \gls{mmwave} solutions to UAV networks is hindered by the severe signal power attenuation experienced at high frequency.
As a matter of fact, \gls{mmwave} propagation  suffers from adverse environmental conditions (principally rain, but also foliage and atmospheric attenuation), human and self blockage, shadowing, and high material penetration loss~\cite{rappaport2014millimeter}. Furthermore,  unlike on-the-ground devices, UAVs are fast-moving, and hence create Doppler. Additionally, the establishment of  directional transmissions  requires periodic beam tracking  to maintain alignment, an operations that may  increase the communication latency, especially in high velocity flight, and result in \gls{qos} degradation~\cite{giordani2018tutorial}.
Moreover,  \gls{uav}  operations are further complicated by high propulsion energy consumption to maintain and support their movements, thereby posing severe power management constraints~\cite{liu2018energy}.  These limitations pose new challenges for proper  protocol design and exemplify how UAV connectivity performance in the \gls{mmwave} scenario is heavily influenced by the specific characteristics of the environment in which the nodes are deployed.

In light of the above challenges, in this paper we apply stochastic geometry to evaluate the practical feasibility of deploying \gls{uav}-based networks operating at \glspl{mmwave}.
To the best of our knowledge, this is the first contribution that provides an analytical expression for the UAV coverage probability for \gls{mmwave} scenarios, i.e., the probability that a reference ground \gls{ue} experiences a link quality (which is measured in terms of \gls{snr}) above a certain threshold.
Our analysis investigates the impact of several UAV-specific parameters on the overall network performance, including the UAV  density, altitude, and  antenna configuration. 
We validate our theoretical model through Monte Carlo simulations and demonstrate that,  while a lower altitude is typically correlated with lower signal power attenuation, it also results in more  likely  non-line-of-sight links due to buildings and other obstacles in the environments, thus leading to intermittent connectivity.
Moreover, we show that a peak in the coverage probability can be associated with an optimal deployment altitude, above which the coverage probability degrades due to the increased \gls{uav}-\gls{ue} distance, and characterize the configurations of the network that minimize the number of deployed \glspl{uav} without compromising the coverage.

The rest of this paper is organized as follows. In Sec.~\ref{sec:related_work} we overview the most recent works on \gls{uav} networks, in Sec.~\ref{sec:stochastic_coverage_analysis} we present our stochastic model for evaluating the coverage probability in a general UAV scenario, in Sec.~\ref{sec:performance_evaluation} we present numerical results and validate our theoretical framework, and in Sec.~\ref{sec:conclusions_and_future_work} we conclude our work with suggestions for future~research.


\section{Related Work} 
\label{sec:related_work}

Given the recent advances in \gls{uav} design and deployment, the research community is studying how cellular and ad hoc networks can benefit from the integration between wireless communications and flying platforms~\cite{zhang2018cellular}. The \gls{3gpp} has examined several study items for the support of UAV-mounted \glspl{ue} in LTE~\cite{36777}, and will consider more advanced solutions for NR, with the possibility of deploying base stations on \glspl{uav} themselves~\cite{22125}. 

The interest in \gls{uav} communications has led to several studies on channel modeling~\cite{al2014optimal,ahmed2016importance}, performance evaluation~\cite{mozaffari2015drone}, and mobility management~\cite{bertizzolo2019mmbac} in flying networks.
Additionally, the combination of \glspl{uav} and mmWaves is seen as a promising enabler of multi-Gbps networks that can be instantiated on the fly~\cite{cuverlier2018mmWaveAerial}, to serve, for example, public safety scenarios~\cite{xia2019millimeter}. A review of the challenges and opportunities for the integration of \glspl{mmwave} and \glspl{uav} can be found in~\cite{xiao2016enabling}. MmWaves can also be used to establish backhaul links in \gls{uav} deployments~\cite{gapeyenko2018flexible}. Two challenges that still  have to be solved in the domain of \glspl{uav} and \glspl{mmwave} are channel modeling (some early results have been presented in~\cite{khawaja2017mmWaveUAVChannel,khawaja2018temporal}) and beam management, which is made more challenging by the high mobility of the flying platform~\cite{zhao2018channel}.

Stochastic geometry has been widely used to characterize the behavior of \gls{uav} networks. In~\cite{ravi2016downlink,chetlur2017downlink} the authors derived the UAV coverage distribution for a network under guaranteed LOS conditions, considering Nakagami-m small-scale fading. The authors of \cite{galkin2017stochastic} model UAV base stations with a \gls{ppp}, varying the height to maximize the coverage probability. Similarly, the authors of \cite{zhang2016spectrum} try to optimize the UAV density, assuming LOS links, while in \cite{azari2017coverage}  a stochastic model jointly optimizes density, height and antenna patterns. Liu \emph{et al.}, in~\cite{liu2018performance}, computed the lower and upper bounds of the coverage probability and area spectral efficiency of UAV networks, assuming, respectively, that UAVs are hovering randomly according to a homogeneous PPP, or that they can instantaneously move to the positions directly above the intended ground users.

In \cite{galkin2018backhaul}, the authors analyze the wireless backhaul links of a UAV network operating in an urban environment in the presence of LOS-blocking buildings, assuming independent distributions of the transmitters and receivers. With respect to the possible applications, \cite{hayajneh2016drone} evaluates the performance of a UAV swarm acting alongside a terrestrial  network in an emergency outage scenario, and \cite{galkin2019stochastic} models the coverage for capacity in hot-spot scenarios. In \cite{zhou2018underlay}, a communication network with an underlay aerial base station is deployed to provide coverage for a temporary event (such as a concert in a stadium or a sporting event). Interestingly, the model divides the space allocated to the temporary event, modeled by a disk of radius $R_2$, and the UAV, deployed at a certain height $h$ at the center of the disk.

While these papers provide solid models and optimizations for UAV wireless networks, they focus on propagation at sub-6 GHz frequencies. The literature currently lacks works combining stochastic geometry models, UAV networks and the \gls{mmwave} band. In this paper we fill this gap by deriving the coverage probability for a stochastic UAV network operating at \glspl{mmwave}, depending on communication and deployment choices (e.g., beamforming design, deployment height and density).

\section{Stochastic Coverage Analysis} 
\label{sec:stochastic_coverage_analysis}

In this section we analyze the coverage of the \gls{uav} scenario based on a stochastic analysis. In detail, in Sec.~\ref{sub:system_model} we describe our system model, in Sec.~\ref{sub:association_rule} we present the association rule for the \gls{ue} and derive a probabilistic expression for the distance between the \gls{ue} and its serving \gls{uav}, while in Sec.~\ref{sub:coverage_probability} we  provide  the expression for the  coverage probability. 

\subsection{System Model} 
\label{sub:system_model}

\begin{figure}[t!]
    \centering
    \setlength\belowcaptionskip{-.3cm}
    \includegraphics[width=0.8\columnwidth]{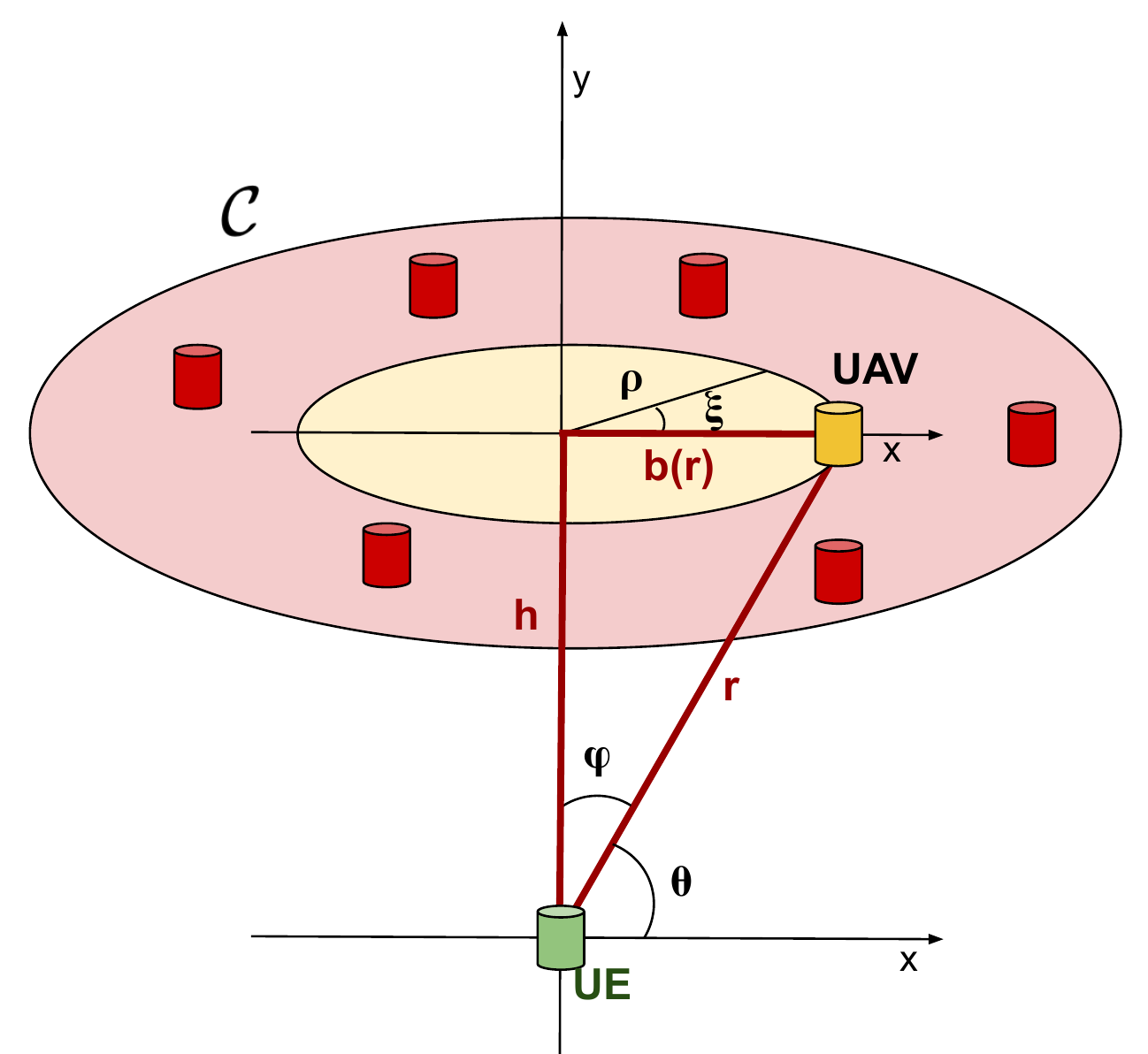}
    \caption{UAV system model. The reference ground \gls{ue} is represented at the origin of a Cartesian system $\{x,y\}$ (green box) and at distance $r$ from the target UAV (yellow box). All other UAVs (red boxes) are deployed in a circular area  according to a \gls{ppp} $\Phi_{\rm UAV}$.}
    \label{fig:model}
\end{figure}



\emph{Network model.} Fig.~\ref{fig:model} provides a graphical representation of the scenario of our analysis.
Without loss of generality, we assume that a reference ground \gls{ue} is placed at the origin of a Cartesian system $\{x,y\}$ centered at point $O=(0,0)$. 
We also assume that  UAVs are deployed across a circular \gls{aoi} $\mathcal{C}$ at an altitude $h$ to form a 2D homogeneous \gls{ppp} $\Phi_{\rm UAV}$ of intensity measure $\lambda_{\rm UAV}$.
We must mention that, even though the network  should  be modeled in the 3D Euclidean space, for the tractability of the analysis, our approach  deploys UAVs at the same altitude $h$ so that the overall 3D system could be seen as a superposition of a first ground  plane $\Pi^{'} = \{x^{'},y^{'}\} \subseteq \mathbb{R}^2$ (with the reference ground \gls{ue} at position $O$) and a second plane $\Pi^{''} = \{x^{''},y^{''}\} \subseteq \mathbb{R}^2$ which includes the AoI $\mathcal{C} \subseteq \mathbb{R}^2$.
We then call $r$ the 2D distance between the reference ground \gls{ue} and a generic UAV in the AoI at altitude $h$, and $b(r) = \sqrt{r^2-h^2}$  the distance between the projection of the \gls{ue} on $\Pi^{''}$ and the generic UAV.

\medskip
\emph{Path loss model.} In the UAV scenario, especially at low elevation angles, we expect that environmental objects (e.g., buildings, vegetation) will act as obstacles to the propagation of the signals. It is thus imperative to distinguish between \gls{los} and \gls{nlos} propagation (denoted with subscripts $L$ and $N$, respectively, throughout the paper). 
In particular, UAV $n$ is assumed to be in \gls{los} (with probability $p_{L}(r)$) if the segment connecting the reference ground \gls{ue} and UAV $n$ (at distance $r$) does not intersect any blockage. In this paper, $p_L(r)$ is modeled as in~\cite{al2014optimal} as a function of  the elevation angle $\theta=\arcsin(h/r)$, i.e., 
\begin{align}\label{plosoverrho}
    p_{L}(r) = \frac{1}{1+C\exp\left[-Y\left(\arcsin\left(\frac{h}{r}\right)\frac{180}{\pi}-C\right)\right]},
\end{align}
where $C$ and $Y$ are environment-related parameters whose values will be detailed in Sec.~\ref{sec:performance_evaluation}.
Similarly, UAVs are assumed to be in \gls{nlos} with complementary probability $p_N(r)=1-p_L(r)$.

By the thinning theorem of PPP~\cite{Bacelli_book}, we can distinguish two independent PPPs for the  LOS and NLOS UAVs, i.e., $\Phi_{\text{UAV},L}\subseteq\Phi_{\rm UAV}$ and $\Phi_{\text{UAV},N}\subseteq\Phi_{\rm UAV}$ respectively, of intensity measure  $\lambda_{\text{UAV},L}=p_L(r)\lambda_{\rm UAV}$, and $\lambda_{\text{UAV},N}=p_N(r)\lambda_{\rm UAV}$ respectively.
Consequently, the path  gain $\ell_i(r)$ between the  reference ground \gls{ue} and the generic UAV $\in\Phi_{\text{UAV},i}$, $i\in\{L,N\}$,  at distance $r$, is expressed as
\begin{equation}
	\label{eq:pathgain}
	\ell_i(r) = C_i r^{-a_i}
\end{equation}
where $a_i$ is the path loss exponent and $C_i$ is the path loss gain at unit distance.

\medskip
\emph{Antenna model.}
As introduced in Sec.~\ref{sec:intro}, \gls{mmwave} communication requires large antenna arrays to be installed at both the UAVs and the UEs to benefit from the resulting antenna gain and overcome the  severe path loss experienced at high frequency.
In this paper, the UAV (UE) antenna array is modeled as a \gls{upa} of $\mathcal{N}_{\rm UAV}$ ($\mathcal{N}_{\rm UE}$) elements. For the tractability of the analysis and consistently with related work on stochastic geometry (e.g.,~\cite{Coverage_rate_analysis,giordani2018coverage}), we assume that antenna patterns are approximated by a sectored antenna model, and we call $G=\mathcal{N}_{\rm UAV}\times\mathcal{N}_{\rm UE}$ the overall antenna gain (assumed constant for all angles in the main lobe) in case of perfect beam alignment between the reference ground UE and its serving~UAV.

\medskip
\emph{Channel model.}
In order to describe correctly the \gls{uav} channel profile at \glspl{mmwave}, small-scale fading is modeled via a Nakagami distribution which is generic enough to incorporate both \gls{los} and \gls{nlos} air-to-ground channels~\cite{zhu2018secrecy} and has been demonstrated to accurately characterize \gls{mmwave} propagation (e.g., reflectivity and scattering from common objects), especially when directional beamforming is applied~\cite{andrews2017modeling}.




\subsection{Association Rule} 
\label{sub:association_rule}

Let $r_n$ be the distance between the reference ground \gls{ue}  and the UAV $n$. We assume that the UE always connects to the UAV $n^* \in \Phi_{\text{UAV},i}$, $i\in\{L,N\}$ that provides the maximum path gain, meaning the UAV whose attributes produce the minimum path loss to the signal, i.e.,
\begin{equation}
    n^* = \argmax_{\forall i \in \{L,N\}, \forall n \in \Phi_i} {\ell_i(r_n)},
\end{equation}
where ${\ell_i(r_n)}$ is given in Eq.~\eqref{eq:pathgain}.

Based on the above definition, we can evaluate the \gls{pdf} of the distance between the reference ground \gls{ue} and the closest available LOS or NLOS UAV (which does not necessarily imply a direct correspondence with the UAV that will be selected for association). 

\begin{lemma}
\label{f_i}
The probability density function of the distance $r$ between the reference ground UE and the closest UAV of type $i = \{L,N\}$ is 
\medmuskip=0mu
\thickmuskip=0mu
    \squeeze_f_i
    \medmuskip=6mu
\thickmuskip=6mu
where  $b(r) = \sqrt{r^2-h^2}$,  $p_i\Big(b(r)\Big)$ is the UAV path loss probability as given in Eq. \eqref{plosoverrho} evaluated at horizontal distance $b(r)$ and $i\in\{L,N\}$ depending on the UAV path loss state.
\end{lemma}

\emph{Proof:} See Appendix \ref{appendix:f_i}.
\hfill $\blacksquare$

\medskip
Another important factor to be taken into account is the possibility of associating not to the closest UAV (distance-wise) but to the one offering the best communication performance, as per the association rule defined previously, especially when considering a very dense urban environment in which the nearest available UAV may be NLOS~\cite{giordani2018coverage}.
Therefore we can model the probability $P_i$ of connecting to a UAV of type $i = \{L,N\}$ as the probability that the distance $r_{i^*}$ between the reference ground UE and the closest UAV of opposite type $i^* = \{N,L\}$ is greater than or equal to $A_i(r)$, defined as
\begin{equation}
    A_i(r) = \left(\frac{C_{i^\ast}}{C_i}r^{a_i}\right)^{\frac{1}{a_i^*}},
\end{equation}
where $i^*$ represents the path loss state opposite to $i$. As a result, assuming that the reference UE connects to a LOS (NLOS) UAV, the problem can be reformulated by modeling the probability that there are no NLOS (LOS) UAVs inside a circular area  of radius $b(A_L(r))$ ($b(A_N(r))$). The related probability $P_i$ is then expressed as in the following lemma.
\begin{lemma}
\label{P_i}
The probability $P_i$ of connecting to a UAV $\in \Phi_{\text{UAV},i}$, with $i = \{L,N\}$, is
\begin{equation}\label{Pi}
    P_i = \int_h^\infty \exp\left(-2\pi\lambda_{\rm UAV}\int_0^{b(A_i(r))} p_{i^\ast}(\rho)\rho d\rho\right)f_i(r) dr,
\end{equation}
with $b(A_i(r)) = \sqrt{A_i(r)^2-h^2}$ being the horizontal distance in the plane $\Pi''\subseteq \mathbb{R}^2$ to the UAV at distance $A_i(r)$ from the reference ground UE.
\end{lemma}

\emph{Proof:} See Appendix \ref{appendix:P_i}.
\hfill $\blacksquare$

\medskip
As a final step, we can derive the expression of the \gls{pdf} of the association distance with a UAV of type $i = \{L,N\}$ in the following lemma.

\begin{lemma}\label{fi_bar_lemma}
The probability density function $\overline{f}_i(r)$ of the association distance to a UAV $\in \Phi_{\text{UAV},i}$, with $i\in\{L,N\}$ at distance $r$ from a reference ground UE is expressed as
\begin{equation}\label{eq:fi_lemma_bar}
    \overline{f}_i(r) = \exp\left(-2\pi\lambda_{UAV}\int_0^{b(A_i(r))}p_{i^\ast}(\rho)) \rho d\rho \right)f_i(r).
\end{equation}
\end{lemma}
\emph{Proof:} The proof follows the same line of reasoning from Lemma~\ref{P_i} and is omitted here.
\hfill $\blacksquare$

\subsection{Coverage Probability} 
\label{sub:coverage_probability}

In this subsection we provide an analytical closed-form expression for the  coverage probability $P_{\rm cov}(\Gamma)$ of a reference ground \gls{ue}, i.e., the probability that the reference user experiences an \gls{snr} larger than a target threshold $\Gamma$ (usually in the order of a few dB). Analytically,  $P_{\rm cov}(\Gamma) = \mathbb{P}[\text{SNR}(r) > \Gamma]$, where the \gls{snr} experienced by the UE, attached to a UAV $\in\Phi_{\rm UAV},i$, with $i\in\{L,N\}$, at distance $r$ is given as
\begin{equation}\label{eq:SNR}
    \text{SNR}_i = \frac{P_{\rm TX}\ell_i(r)Gg}{\text{NF}\cdot\sigma^2}.
\end{equation} 
In Eq.~\eqref{eq:SNR}, $P_{\rm TX}$ is the transmit power, $\ell_i(r)$ is the path gain profile as expressed in Eq.~\eqref{eq:pathgain}, $G = \mathcal{N}_{\rm UAV}\times \mathcal{N}_{\rm UE}$ is the aggregate beamforming gain, $g$ is the small-scale fading, NF is the noise figure and $\sigma^2$ is the power of the thermal noise.

By the law of total probability, we can split the SNR contribution due to the presence of LOS and NLOS conditions for the channel, so the coverage probability becomes
\medmuskip=4mu
\thickmuskip=4mu
\begin{align}\label{eq:SNR_divided}
    &P_{\rm cov}(\Gamma) = \mathbb{P}\Big[\text{SNR} > \Gamma \Big] \notag \\
    &= \mathbb{P}\Big[\text{SNR}_L > \Gamma, n^* \in \Phi_L\Big] + \mathbb{P}\Big[\text{SNR}_N > \Gamma, n^* \in \Phi_N\Big],
\end{align}

    \medmuskip=6mu
\thickmuskip=6mu
where $n^*$ is the serving UAV according to the association rule defined in Sec.~\ref{sub:association_rule}.
Based on the lemmas and assumptions of the previous sections, we can formalize the SNR coverage probability in the following theorem. 
\begin{theorem}\label{th:pcov}
The coverage probability $P_{\rm cov}(\Gamma)$ for an SNR threshold $\Gamma$, considering a Nakagami$(m_i,\Omega_i)$ distributed fading $g$ of shape $m_i$ and spread $\Omega_i$, with $i\in\{L,N\}$, is given~by
\begin{align}\label{Pcov_value_theorem}
    P_{\rm cov}(\Gamma) &= 
    \sum_{i\in\{L,N\}}\int_h^{\infty} \left(1-\frac{\int_0^{\Delta_i}t^{m_i-1}e^{-t}dt}{\left(m_i-1\right)!}\right)\overline{f}_i(r)dr,
\end{align}
where $\Delta_i = \frac{m_i}{\Omega_i}\zeta_i(r)^2$ and $\zeta_i(r)=\frac{\Gamma\left(\text{NF}\cdot\sigma^2\right)}{P_{\rm TX}GC_i}r^{a_i}$.
\end{theorem}
\emph{Proof:} See Appendix \ref{appendix:P_cov}.
\hfill $\blacksquare$

\section{Performance Evaluation} 
\label{sec:performance_evaluation}

This section reports a numerical evaluation of the coverage for \glspl{uav} in mmWave scenarios. The results are based on the analytical model introduced in Sec.~\ref{sec:stochastic_coverage_analysis}, and on a Monte Carlo simulation campaign that we run to validate our analysis.

\subsection{Scenario}

We deploy \glspl{uav} operating at $f_c=28$ GHz. For the channel model, we consider the parameters reported in Table~\ref{table:channelparam}: the \gls{los} probability for \glspl{uav} follows Eq.~\eqref{plosoverrho}, the path gain is modeled according to~\cite{Mustafa}, which considers measurements in a typical urban context (i.e., New York City), while the fading is modeled with a Nakagami random variable with parameters from~\cite{zhu2018secrecy}.

Table~\ref{table:netparam} summarizes the configuration parameters for the network. Notably, in the following, we will evaluate the coverage by varying (i) the number of antenna elements $\mathcal{N}_{\rm UAV}$ and $\mathcal{N}_{\rm UE}$ in the arrays of the \glspl{uav} and \glspl{ue}, respectively; (ii) the SNR threshold $\Gamma$, which discriminates a link with a high enough quality or not; (iii) the altitude $h$ at which the \glspl{uav} are deployed; and (iv) the density $\lambda_{\rm UAV}$ in terms of \gls{uav}/km$^2$.

For the Monte Carlo simulation, we generated $1000$ random realizations of a \gls{ppp} in an AoI of radius $2000$ m, for each configuration of parameters. In the following figures, the markers indicate the Monte Carlo simulation results, while the lines represent the numerical results for the analytical model, solved using the MATLAB Symbolic~toolbox.

\begin{table}[t]
	\centering
    \caption{Parameters of the channel model, from~\cite{Mustafa,zhu2018secrecy,al2014optimal}.}
  \label{table:channelparam}
 	\begin{tabular}{lll}
 	\toprule
 	Parameter & LOS Value & NLOS Value \\\midrule
 	Nakagami-m shape parameter & $m_L = 3$ & $m_N = 2$  \\
 	Nakagami-m spread parameter & $\Omega_L = 1$ & $\Omega_N = 1$  \\
 	Path gain intercept & $C_L = 10^{-6.14}$ & $C_N = 10^{-7.2}$ \\
 	Path gain exponent & $a_L = 2$ & $a_N = 2.92$ \\
 	\bottomrule
 	\toprule
 	Urban LOS probability parameters & $C=9.6117$ & $Y=0.1581$ \\ 
 	\bottomrule
 	\end{tabular}
 \end{table}

\begin{table}[t]
	\centering
    \caption{Parameters of the communication scenario}
  \label{table:netparam}
 	\begin{tabular}{ll}
 	\toprule
 	Parameter & Value \\\midrule
 	Transmit power $P_{\rm TX}$ & 20 dBm\\
 	Antenna configurations ($\mathcal{N}_{\rm UAV} \times \mathcal{N}_{\rm UE}$) & $[8\times4]$, $[8\times8]$, $[64\times4]$\\
	&$[256\times4]$, $[256\times8]$\\
 	Noise figure NF & 5 dB \\
 	Thermal noise $\sigma^2$ & -84 dBm\\
 	SNR threshold $\Gamma$ & $[-5, 0, 5]$ dB \\
 	UAV deployment height $h$ & $[0-1000]$ m \\
 	UAV deployment density $\lambda_{\rm UAV}$ & $[1, 5, 10, 15, 25]$ UAVs/km$^2$ \\
 	\bottomrule
 	\end{tabular}
 	\vspace{-0.33cm}
 \end{table} 


\begin{figure*}
	\begin{subfigure}[t]{0.32\textwidth}
		\centering
		\setlength\fwidth{0.8\textwidth}
		\setlength\fheight{0.5\textwidth}
%
%
\definecolor{mycolor1}{rgb}{0.00000,0.44700,0.74100}%
\definecolor{mycolor2}{rgb}{0.92900,0.69400,0.12500}%
\definecolor{mycolor3}{rgb}{0.46600,0.67400,0.18800}%
\definecolor{mycolor4}{rgb}{0.63500,0.07800,0.18400}%
\definecolor{mycolor5}{rgb}{0.85000,0.32500,0.09800}%
\begin{tikzpicture}
\pgfplotsset{every tick label/.append style={font=\scriptsize}}

\begin{axis}[%
width=0.951\fwidth,
height=\fheight,
at={(0\fwidth,0\fheight)},
scale only axis,
xmin=0,
xmax=1000,
xlabel style={font=\footnotesize\color{white!15!black}},
xlabel={Altitude $h$ [m]},
ymin=0,
ymax=1,
ylabel style={font=\footnotesize\color{white!15!black}},
ylabel={$P_{\rm cov}$},
axis background/.style={fill=white},
xmajorgrids,
ymajorgrids,
legend columns=5,
legend style={at={(-.15,1.05)}, anchor=south west, legend cell align=left, align=left, draw=white!15!black, font=\footnotesize}
]
\addplot [color=mycolor1]
  table[row sep=crcr]{%
0	0.0820384562819114\\
50	0.191585137503984\\
100	0.357415368596249\\
150	0.518192902728245\\
200	0.648840762296214\\
250	0.743586333020858\\
300	0.806594382517935\\
350	0.845084270367822\\
400	0.865723857816848\\
450	0.873381590037471\\
500	0.871063246357289\\
550	0.860247122228182\\
600	0.841256932145222\\
650	0.813581684231634\\
700	0.776179746629115\\
750	0.727844095849231\\
800	0.667687795684475\\
850	0.595740862049408\\
900	0.513542331694604\\
950	0.424503826752418\\
1000	0.333786223895937\\
};
\addlegendentry{$\lambda_{\rm UAV} = 1$ UAVs/km$^2$}

\addplot [color=mycolor1, line width=1.0pt, draw=none, mark size=1.3pt, mark=triangle*, mark options={solid, fill=mycolor1, mycolor1}, forget plot]
  table[row sep=crcr]{%
0	0.081\\
50	0.176\\
100	0.336\\
150	0.505\\
200	0.66\\
250	0.747\\
300	0.833\\
350	0.843\\
400	0.864\\
450	0.879\\
500	0.868\\
550	0.869\\
600	0.824\\
650	0.81\\
700	0.784\\
750	0.687\\
800	0.681\\
850	0.594\\
900	0.526\\
950	0.424\\
1000	0.34\\
};

\addplot [color=mycolor2]
  table[row sep=crcr]{%
0	0.344164219969877\\
50	0.650480329878782\\
100	0.887216680450047\\
150	0.972147259968835\\
200	0.993641127652316\\
250	0.998275843758074\\
300	0.999231742061137\\
350	0.99934237992656\\
400	0.999110819172156\\
450	0.998498478872289\\
500	0.99720562954669\\
550	0.994612866420564\\
600	0.98961141248367\\
650	0.980389769215693\\
700	0.96425986034513\\
750	0.937661284706159\\
800	0.896519423849437\\
850	0.837090534491853\\
900	0.757253792324343\\
950	0.657920351621631\\
1000	0.543951062045194\\
};
\addlegendentry{$\lambda_{\rm UAV} = 5$ UAVs/km$^2$}

\addplot [color=mycolor2, line width=1.0pt, draw=none, mark size=1.3pt, mark=triangle*, mark options={solid, fill=mycolor2, mycolor2}, forget plot]
  table[row sep=crcr]{%
0	0.368\\
50	0.639\\
100	0.891\\
150	0.974\\
200	0.993\\
250	1\\
300	0.999\\
350	0.998\\
400	0.999\\
450	1\\
500	0.999\\
550	0.994\\
600	0.99\\
650	0.982\\
700	0.965\\
750	0.937\\
800	0.893\\
850	0.844\\
900	0.759\\
950	0.653\\
1000	0.528\\
};

\addplot [color=mycolor3]
  table[row sep=crcr]{%
0	0.563359319699954\\
50	0.87410811199029\\
100	0.986312949726569\\
150	0.999060981526832\\
200	0.999927735163262\\
250	0.999980669843528\\
300	0.999975477965551\\
350	0.99994524287677\\
400	0.999858079062107\\
450	0.999622914745062\\
500	0.99902464270455\\
550	0.997602309576067\\
600	0.994458206117505\\
650	0.988008627469649\\
700	0.975739241327414\\
750	0.954102041775562\\
800	0.918753436731041\\
850	0.865320274906835\\
900	0.790724104871317\\
950	0.694783110053158\\
1000	0.581462349615432\\
};
\addlegendentry{$\lambda_{\rm UAV} = 10$ UAVs/km$^2$}

\addplot [color=mycolor3, line width=1.0pt, draw=none, mark size=1.3pt, mark=triangle*, mark options={solid, fill=mycolor3, mycolor3}, forget plot]
  table[row sep=crcr]{%
0	0.589\\
50	0.874\\
100	0.985\\
150	0.999\\
200	1\\
250	1\\
300	1\\
350	1\\
400	1\\
450	1\\
500	1\\
550	1\\
600	0.991\\
650	0.992\\
700	0.976\\
750	0.948\\
800	0.912\\
850	0.86\\
900	0.778\\
950	0.685\\
1000	0.6\\
};

\addplot [color=mycolor4]
  table[row sep=crcr]{%
0	0.705052773321287\\
50	0.953308216326364\\
100	0.9982101741886\\
150	0.999960171977244\\
200	0.999997562084999\\
250	0.999997475405253\\
300	0.999992935251661\\
350	0.999976584201996\\
400	0.999922417853093\\
450	0.999757545198484\\
500	0.999299826425123\\
550	0.998141041989435\\
600	0.99545820511049\\
650	0.98975839494174\\
700	0.978614380354749\\
750	0.958525081139516\\
800	0.925105613042588\\
850	0.873811052395306\\
900	0.801251751701942\\
950	0.706845256662621\\
1000	0.594179084734891\\
};
\addlegendentry{$\lambda_{\rm UAV} = 15$ UAVs/km$^2$}

\addplot [color=mycolor4, line width=1.0pt, draw=none, mark size=1.3pt, mark=triangle*, mark options={solid, fill=mycolor4, mycolor4}, forget plot]
  table[row sep=crcr]{%
0	0.705\\
50	0.946\\
100	1\\
150	1\\
200	1\\
250	1\\
300	1\\
350	1\\
400	1\\
450	1\\
500	1\\
550	0.996\\
600	0.999\\
650	0.988\\
700	0.973\\
750	0.947\\
800	0.91\\
850	0.893\\
900	0.804\\
950	0.7\\
1000	0.589\\
};

\addplot [color=mycolor5]
  table[row sep=crcr]{%
0	0.859747547774941\\
50	0.99301038767277\\
100	0.999961432557519\\
150	0.999999786846017\\
200	0.999999858932529\\
250	0.999999415909338\\
300	0.99999719935817\\
350	0.999987438981019\\
400	0.999949987442867\\
450	0.999824335467715\\
500	0.999451632408463\\
550	0.998462868888267\\
600	0.996093417173684\\
650	0.990925266996053\\
700	0.980608968379018\\
750	0.961695886326181\\
800	0.929788111200294\\
850	0.880223033717041\\
900	0.809373182252371\\
950	0.716330096240352\\
1000	0.604354256750157\\
};
\addlegendentry{$\lambda_{\rm UAV} = 25$ UAVs/km$^2$}

\addplot [color=mycolor5, line width=1.0pt, draw=none, mark size=1.3pt, mark=triangle*, mark options={solid, fill=mycolor5, mycolor5}, forget plot]
  table[row sep=crcr]{%
0	0.848\\
50	0.997\\
100	1\\
150	1\\
200	1\\
250	1\\
300	1\\
350	1\\
400	1\\
450	0.999\\
500	0.999\\
550	0.999\\
600	0.996\\
650	0.99\\
700	0.987\\
750	0.958\\
800	0.907\\
850	0.881\\
900	0.802\\
950	0.707\\
1000	0.609\\
};

\node[align=center] (coord) at (500, 0.2) {\scriptsize$h_{\rm opt}=338$ m \\\scriptsize for $\lambda_{\rm UAV} = 5$ UAVs/km$^2$};
\draw[->] (coord) -- (338, 0.994);

\end{axis}

\end{tikzpicture}%
		\caption{$\Gamma = -5$ dB}
		\label{fig:minus5db}
	\end{subfigure}\hfill
	\begin{subfigure}[t]{0.32\textwidth}
		\centering
		\setlength\fwidth{0.8\textwidth}
		\setlength\fheight{0.5\textwidth}
%
%
\definecolor{mycolor1}{rgb}{0.00000,0.44700,0.74100}%
\definecolor{mycolor2}{rgb}{0.92900,0.69400,0.12500}%
\definecolor{mycolor3}{rgb}{0.46600,0.67400,0.18800}%
\definecolor{mycolor4}{rgb}{0.63500,0.07800,0.18400}%
\definecolor{mycolor5}{rgb}{0.85000,0.32500,0.09800}%
\begin{tikzpicture}
\pgfplotsset{every tick label/.append style={font=\scriptsize}}

\begin{axis}[%
width=0.951\fwidth,
height=\fheight,
at={(0\fwidth,0\fheight)},
scale only axis,
xmin=0,
xmax=1000,
xlabel style={font=\footnotesize\color{white!15!black}},
xlabel={Altitude $h$ [m]},
ymin=0,
ymax=1,
ylabel style={font=\footnotesize\color{white!15!black}},
ylabel={$P_{\rm cov}$},
axis background/.style={fill=white},
xmajorgrids,
ymajorgrids,
legend columns=5,
legend style={font=\footnotesize, legend cell align=left, align=left, draw=white!15!black, anchor=south, at={(1.2,1.05)}}
]
\addplot [color=mycolor1]
  table[row sep=crcr]{%
0	0.0278167222842979\\
50	0.114920976620632\\
100	0.24967780262429\\
150	0.372991214294411\\
200	0.458111243791103\\
250	0.498982176139837\\
300	0.501079319271536\\
350	0.472739660804109\\
400	0.420432085147835\\
450	0.348777169732946\\
500	0.263682204140653\\
550	0.1755823060118\\
600	0.0987170301734631\\
650	0.0446870013636397\\
700	0.0154850625415969\\
750	0.00389829049930508\\
800	0.000675646986135944\\
850	7.62788701103956e-05\\
900	5.2970261290083e-06\\
950	2.13159150697704e-07\\
1000	4.67075002121175e-09\\
};

\addplot [color=mycolor1, line width=1.0pt, draw=none, mark size=1.3pt, mark=triangle*, mark options={solid, fill=mycolor1, mycolor1}]
  table[row sep=crcr]{%
0	0.031\\
50	0.106\\
100	0.231\\
150	0.36\\
200	0.452\\
250	0.518\\
300	0.52\\
350	0.476\\
400	0.433\\
450	0.36\\
500	0.261\\
550	0.193\\
600	0.099\\
650	0.053\\
700	0.016\\
750	0.002\\
800	0\\
850	0\\
900	0\\
950	0\\
1000	0\\
};

\addplot [color=mycolor2]
  table[row sep=crcr]{%
0	0.130979464008721\\
50	0.45556376419224\\
100	0.759571804662879\\
150	0.89841732955693\\
200	0.945149520418434\\
250	0.954404319961341\\
300	0.944827574142495\\
350	0.917204255824838\\
400	0.863314028424327\\
450	0.770762465047793\\
500	0.63170575892295\\
550	0.455879254030196\\
600	0.276271003846897\\
650	0.133719795983387\\
700	0.0491135373244024\\
750	0.0129970896772076\\
800	0.00235078599957935\\
850	0.000275253993271182\\
900	1.97222573911818e-05\\
950	8.15388381543706e-07\\
1000	1.82911131627144e-08\\
};

\addplot [color=mycolor2, line width=1.0pt, draw=none, mark size=1.3pt, mark=triangle*, mark options={solid, fill=mycolor2, mycolor2}]
  table[row sep=crcr]{%
0	0.14\\
50	0.446\\
100	0.743\\
150	0.894\\
200	0.943\\
250	0.96\\
300	0.946\\
350	0.91\\
400	0.846\\
450	0.75\\
500	0.629\\
550	0.42\\
600	0.284\\
650	0.141\\
700	0.044\\
750	0.01\\
800	0\\
850	0\\
900	0\\
950	0\\
1000	0\\
};

\addplot [color=mycolor3]
  table[row sep=crcr]{%
0	0.243553402529056\\
50	0.701813421619297\\
100	0.94052611716028\\
150	0.98816890606077\\
200	0.995045420259482\\
250	0.994350566462396\\
300	0.988997875297005\\
350	0.974337543235144\\
400	0.939130006776147\\
450	0.865505370317775\\
500	0.7364700364716\\
550	0.553269916385851\\
600	0.34896750640715\\
650	0.175421708435085\\
700	0.0667090362282581\\
750	0.018216543051926\\
800	0.0033889044070365\\
850	0.000406929241952123\\
900	2.98220370254177e-05\\
950	1.25816253582556e-06\\
1000	2.87426027934513e-08\\
};

\addplot [color=mycolor3, line width=1.0pt, draw=none, mark size=1.3pt, mark=triangle*, mark options={solid, fill=mycolor3, mycolor3}]
  table[row sep=crcr]{%
0	0.251\\
50	0.724\\
100	0.943\\
150	0.989\\
200	1\\
250	0.993\\
300	0.992\\
350	0.974\\
400	0.945\\
450	0.892\\
500	0.727\\
550	0.569\\
600	0.31\\
650	0.188\\
700	0.067\\
750	0.021\\
800	0.004\\
850	0\\
900	0\\
950	0\\
1000	0\\
};

\addplot [color=mycolor4]
  table[row sep=crcr]{%
0	0.340467368681009\\
50	0.835704141970833\\
100	0.984830566420579\\
150	0.998348452990236\\
200	0.999159891290942\\
250	0.998369282584016\\
300	0.995243318210201\\
350	0.985277171319173\\
400	0.957672724947259\\
450	0.893625312881783\\
500	0.772691367569048\\
550	0.591286141479772\\
600	0.38026546790534\\
650	0.194882468776511\\
700	0.0754940779818833\\
750	0.0209770451444046\\
800	0.00396603129252011\\
850	0.000483404186798499\\
900	3.59195046294523e-05\\
950	1.53488754335147e-06\\
1000	3.54810981491979e-08\\
};

\addplot [color=mycolor4, line width=1.0pt, draw=none, mark size=1.3pt, mark=triangle*, mark options={solid, fill=mycolor4, mycolor4}]
  table[row sep=crcr]{%
0	0.351\\
50	0.848\\
100	0.986\\
150	0.999\\
200	1\\
250	0.998\\
300	0.995\\
350	0.979\\
400	0.967\\
450	0.899\\
500	0.764\\
550	0.565\\
600	0.376\\
650	0.201\\
700	0.069\\
750	0.021\\
800	0.004\\
850	0.001\\
900	0\\
950	0\\
1000	0\\
};

\addplot [color=mycolor5]
  table[row sep=crcr]{%
0	0.496211488507622\\
50	0.949218070247852\\
100	0.998905343721597\\
150	0.999927965307999\\
200	0.999882026216246\\
250	0.999503729873927\\
300	0.997741675697474\\
350	0.990851987251886\\
400	0.968937206302956\\
450	0.913109698628655\\
500	0.800470581267516\\
550	0.622897024981745\\
600	0.40807758398941\\
650	0.213174338656696\\
700	0.0841625479118925\\
750	0.0238202394791961\\
800	0.00458372764523337\\
850	0.000568156880593136\\
900	4.28958824523776e-05\\
950	1.86093991526634e-06\\
1000	4.36397731174603e-08\\
};

\addplot [color=mycolor5, line width=1.0pt, draw=none, mark size=1.3pt, mark=triangle*, mark options={solid, fill=mycolor5, mycolor5}]
  table[row sep=crcr]{%
0	0.486\\
50	0.955\\
100	0.999\\
150	1\\
200	1\\
250	1\\
300	0.998\\
350	0.994\\
400	0.959\\
450	0.914\\
500	0.795\\
550	0.655\\
600	0.381\\
650	0.236\\
700	0.09\\
750	0.023\\
800	0.008\\
850	0\\
900	0\\
950	0\\
1000	0\\
};

\node[align=right] (coord) at (760, 0.7) {\scriptsize $h_{\rm opt}=246$ m\\\scriptsize for $\lambda_{\rm UAV} =$\\\scriptsize $5$ UAVs/km$^2$};
\draw[->] (coord) -- (246, 0.954453522793302);

\end{axis}
\end{tikzpicture}%
		\caption{$\Gamma = 0$ dB}
		\label{fig:0db}
	\end{subfigure}\hfill
	\begin{subfigure}[t]{0.32\textwidth}
		\centering
		\setlength\fwidth{0.8\textwidth}
		\setlength\fheight{0.5\textwidth}
%
%
\definecolor{mycolor1}{rgb}{0.00000,0.44700,0.74100}%
\definecolor{mycolor2}{rgb}{0.92900,0.69400,0.12500}%
\definecolor{mycolor3}{rgb}{0.46600,0.67400,0.18800}%
\definecolor{mycolor4}{rgb}{0.63500,0.07800,0.18400}%
\definecolor{mycolor5}{rgb}{0.85000,0.32500,0.09800}%
\begin{tikzpicture}
\pgfplotsset{every tick label/.append style={font=\scriptsize}}

\begin{axis}[%
width=0.951\fwidth,
height=\fheight,
at={(0\fwidth,0\fheight)},
scale only axis,
xmin=0,
xmax=1000,
xlabel style={font=\footnotesize\color{white!15!black}},
xlabel={Altitude $h$ [m]},
ymin=0,
ymax=1,
ylabel style={font=\footnotesize\color{white!15!black}},
ylabel={$P_{\rm cov}$},
axis background/.style={fill=white},
xmajorgrids,
ymajorgrids,
legend style={legend cell align=left, align=left, draw=white!15!black}
]
\addplot [color=mycolor1]
  table[row sep=crcr]{%
0	0.0093731924558771\\
50	0.0753949695510374\\
100	0.161725417745631\\
150	0.202204359921727\\
200	0.187756612719733\\
250	0.138668184426623\\
300	0.0759584713231941\\
350	0.0255560533709528\\
400	0.0040677361542137\\
450	0.000228525650679477\\
500	3.32271473646654e-06\\
550	8.99275092986677e-09\\
600	3.17885984256872e-12\\
650	1.00067424261626e-16\\
700	1.85412108879398e-22\\
750	1.29413916600076e-29\\
800	2.10729089220403e-38\\
850	4.79595920122853e-49\\
900	8.84071952940258e-62\\
950	7.39841580003657e-77\\
1000	1.52378972487226e-94\\
};

\addplot [color=mycolor1, line width=1.0pt, draw=none, mark size=1.3pt, mark=triangle*, mark options={solid, fill=mycolor1, mycolor1}]
  table[row sep=crcr]{%
0	0.008\\
50	0.066\\
100	0.149\\
150	0.178\\
200	0.194\\
250	0.148\\
300	0.072\\
350	0.031\\
400	0.003\\
450	0\\
500	0\\
550	0\\
600	0\\
650	0\\
700	0\\
750	0\\
800	0\\
850	0\\
900	0\\
950	0\\
1000	0\\
};

\addplot [color=mycolor2]
  table[row sep=crcr]{%
0	0.0459254624758646\\
50	0.323735592089653\\
100	0.582369457712369\\
150	0.663101140268572\\
200	0.617998709062204\\
250	0.483799271790203\\
300	0.284986840703404\\
350	0.102282757038332\\
400	0.0171079232934975\\
450	0.000996420601449275\\
500	1.48715990349133e-05\\
550	4.10341178118605e-08\\
600	1.47187742257082e-11\\
650	4.6861033470602e-16\\
700	8.76092124222347e-22\\
750	6.15926919847557e-29\\
800	1.00887981587139e-37\\
850	2.30737945390568e-48\\
900	4.27087537477391e-61\\
950	3.5865869297766e-76\\
1000	7.40903539771433e-94\\
};

\addplot [color=mycolor2, line width=1.0pt, draw=none, mark size=1.3pt, mark=triangle*, mark options={solid, fill=mycolor2, mycolor2}]
  table[row sep=crcr]{%
0	0.042\\
50	0.321\\
100	0.582\\
150	0.672\\
200	0.605\\
250	0.514\\
300	0.277\\
350	0.104\\
400	0.017\\
450	0.001\\
500	0\\
550	0\\
600	0\\
650	0\\
700	0\\
750	0\\
800	0\\
850	0\\
900	0\\
950	0\\
1000	0\\
};

\addplot [color=mycolor3]
  table[row sep=crcr]{%
0	0.0895748582821106\\
50	0.541775911863377\\
100	0.82142093533871\\
150	0.873376375398253\\
200	0.824339349383409\\
250	0.679954553248326\\
300	0.426895820164987\\
350	0.162377038977363\\
400	0.0284411733713782\\
450	0.00171513070664499\\
500	2.6275437404963e-05\\
550	7.39559409882382e-08\\
600	2.69406995293033e-11\\
650	8.68288241633336e-16\\
700	1.63940279891181e-21\\
750	1.1619039649175e-28\\
800	1.91595809500382e-37\\
850	4.40657774074202e-48\\
900	8.19524119281078e-61\\
950	6.91015598739236e-76\\
1000	1.43246737561369e-93\\
};

\addplot [color=mycolor3, line width=1.0pt, draw=none, mark size=1.3pt, mark=triangle*, mark options={solid, fill=mycolor3, mycolor3}]
  table[row sep=crcr]{%
0	0.11\\
50	0.559\\
100	0.836\\
150	0.874\\
200	0.827\\
250	0.664\\
300	0.424\\
350	0.147\\
400	0.032\\
450	0.003\\
500	0\\
550	0\\
600	0\\
650	0\\
700	0\\
750	0\\
800	0\\
850	0\\
900	0\\
950	0\\
1000	0\\
};

\addplot [color=mycolor4]
  table[row sep=crcr]{%
0	0.1310685539131\\
50	0.688903139001497\\
100	0.921616250017566\\
150	0.946551451052675\\
200	0.904813232171474\\
250	0.772861103708001\\
300	0.506807106710991\\
350	0.200907785073381\\
400	0.0364126694214969\\
450	0.00225535911767782\\
500	3.52767371927347e-05\\
550	1.00923041913671e-07\\
600	3.72450188966187e-11\\
650	1.21308222812194e-15\\
700	2.31025860963771e-21\\
750	1.64914852092476e-28\\
800	2.73584249281859e-37\\
850	6.32446289960968e-48\\
900	1.18135128790101e-60\\
950	9.99855321860745e-76\\
1000	2.07945237402525e-93\\
};

\addplot [color=mycolor4, line width=1.0pt, draw=none, mark size=1.3pt, mark=triangle*, mark options={solid, fill=mycolor4, mycolor4}]
  table[row sep=crcr]{%
0	0.137\\
50	0.717\\
100	0.924\\
150	0.94\\
200	0.894\\
250	0.778\\
300	0.505\\
350	0.203\\
400	0.032\\
450	0.002\\
500	0\\
550	0\\
600	0\\
650	0\\
700	0\\
750	0\\
800	0\\
850	0\\
900	0\\
950	0\\
1000	0\\
};

\addplot [color=mycolor5]
  table[row sep=crcr]{%
0	0.208037020258117\\
50	0.855739106713979\\
100	0.983462694324809\\
150	0.986520817359597\\
200	0.959141491554811\\
250	0.852304336324459\\
300	0.589158822700856\\
350	0.24635555868691\\
400	0.0467729094612626\\
450	0.00300950540273516\\
500	4.85452594339873e-05\\
550	1.42406876276774e-07\\
600	5.36481788145109e-11\\
650	1.77751592133499e-15\\
700	3.4342210384722e-21\\
750	2.48152088275482e-28\\
800	4.15975005362867e-37\\
850	9.70253993531933e-48\\
900	1.82643450175815e-60\\
950	1.55628939458531e-75\\
1000	3.25585658544824e-93\\
};

\addplot [color=mycolor5, line width=1.0pt, draw=none, mark size=1.3pt, mark=triangle*, mark options={solid, fill=mycolor5, mycolor5}]
  table[row sep=crcr]{%
0	0.199\\
50	0.873\\
100	0.987\\
150	0.99\\
200	0.951\\
250	0.862\\
300	0.582\\
350	0.244\\
400	0.049\\
450	0.001\\
500	0\\
550	0\\
600	0\\
650	0\\
700	0\\
750	0\\
800	0\\
850	0\\
900	0\\
950	0\\
1000	0\\
};

\end{axis}

\begin{axis}[%
width=0.951\fwidth,
height=\fheight,
at={(0\fwidth,0\fheight)},
scale only axis,
xmin=0,
xmax=1000,
xlabel style={font=\footnotesize\color{white!15!black}},
xlabel={Altitude $h$ [m]},
ymin=0,
ymax=1,
ylabel style={font=\footnotesize\color{white!15!black}},
ylabel={$P_{\rm cov}$},
xmajorgrids,
ymajorgrids,
legend style={legend cell align=left, align=left, draw=white!15!black, font=\footnotesize, at={(0.98, 1.06)}, anchor=north east},
hide y axis,
hide x axis,
]

\addplot [color=black]
  table[row sep=crcr]{%
-1 -1\\
};
\addlegendentry{Analysis}

\addplot [color=black, draw=none, mark size=1.3pt, mark=triangle*, mark options={solid, fill=black, black}, only marks]
  table[row sep=crcr]{%
-1 -1\\
};
\addlegendentry{Monte Carlo}

\node[align=left] (coord) at (700, 0.4) {\scriptsize $h_{\rm opt}=153$ m for\\\scriptsize $\lambda_{\rm UAV} = 5$ UAVs/km$^2$};
\draw[->] (coord) -- (153, 0.663355880663136);

\end{axis}

\end{tikzpicture}%
		\caption{$\Gamma = 5$ dB}
		\label{fig:5db}
	\end{subfigure}
  \setlength\belowcaptionskip{-.3cm}
	\caption{Coverage probability $P_{\rm cov}$ vs. the deployment height $h$ and the SNR threshold $\Gamma$. The antenna configuration has $\mathcal{N}_{\rm UAV} = 8$ and $\mathcal{N}_{\rm UE} = 8$. The lines  represent the numerical results given by the analytical model, and the markers the Monte Carlo~simulations.}
	\label{fig:snr_thresh}
\end{figure*}
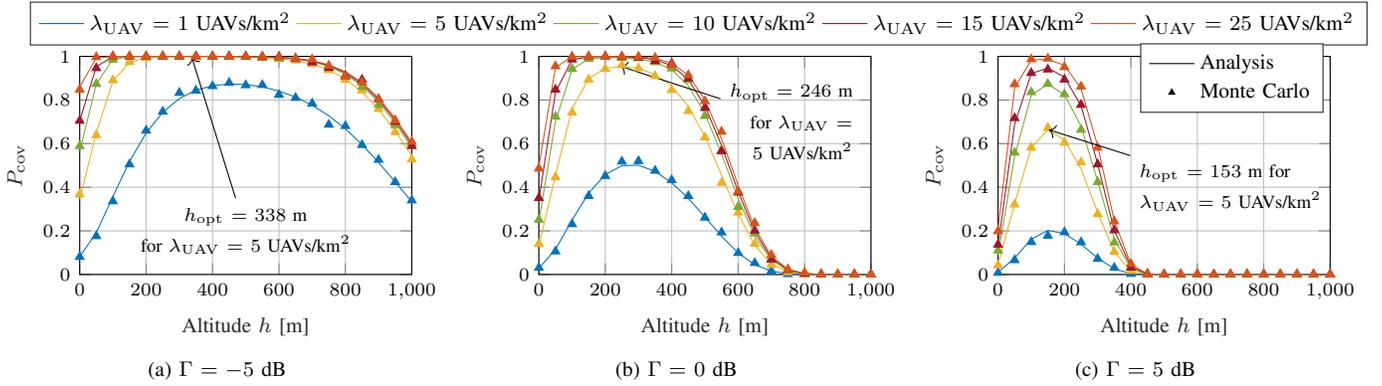

\subsection{Impact of the SNR threshold}

\begin{figure}[t]
    \centering
    \setlength\fwidth{0.7\columnwidth}
    \setlength\fheight{0.35\columnwidth}
%
%
\definecolor{mycolor1}{rgb}{0.00000,0.44700,0.74100}%
\definecolor{mycolor2}{rgb}{0.85000,0.32500,0.09800}%
\definecolor{mycolor3}{rgb}{0.92900,0.69400,0.12500}%
\definecolor{mycolor4}{rgb}{0.49400,0.18400,0.55600}%
\definecolor{mycolor5}{rgb}{0.46600,0.67400,0.18800}%
\begin{tikzpicture}
\pgfplotsset{every tick label/.append style={font=\scriptsize}}

\begin{axis}[%
width=0.951\fwidth,
height=\fheight,
at={(0\fwidth,0\fheight)},
scale only axis,
xmin=0,
xmax=25,
xlabel style={font=\footnotesize\color{white!15!black}},
xlabel={UAV density $\lambda_{\rm UAV}$ [UAVs/km$^2$]},
ymin=0,
ymax=1,
ylabel style={font=\footnotesize\color{white!15!black}},
ylabel={$P_{\rm cov}$},
axis background/.style={fill=white},
xmajorgrids,
ymajorgrids,
legend style={at={(0.5, 1.05)}, anchor=south, legend cell align=left, align=left, draw=white!15!black, font=\footnotesize},
legend columns=3
]
\addplot [color=mycolor1, draw=none, mark size=1.3pt, mark=triangle*, mark options={solid, fill=mycolor1, mycolor1}, forget plot]
  table[row sep=crcr]{%
1	0.047\\
5	0.212\\
10	0.362\\
15	0.446\\
25	0.58\\
};

\addplot [color=mycolor2, draw=none, mark size=1.3pt, mark=triangle*, mark options={solid, fill=mycolor2, mycolor2}, forget plot]
  table[row sep=crcr]{%
1	0.163\\
5	0.617\\
10	0.836\\
15	0.905\\
25	0.962\\
};

\addplot [color=mycolor3, draw=none, mark size=1.3pt, mark=triangle*, mark options={solid, fill=mycolor3, mycolor3}, forget plot]
  table[row sep=crcr]{%
1	0.493\\
5	0.964\\
10	1\\
15	1\\
25	1\\
};

\addplot [color=mycolor4, draw=none, mark size=1.3pt, mark=triangle*, mark options={solid, fill=mycolor4, mycolor4}, forget plot]
  table[row sep=crcr]{%
1	0.698\\
5	0.997\\
10	1\\
15	1\\
25	1\\
};

\addplot [color=mycolor5, draw=none, mark size=1.3pt, mark=triangle*, mark options={solid, fill=mycolor5, mycolor5}, forget plot]
  table[row sep=crcr]{%
1	0.8\\
5	0.999\\
10	1\\
15	1\\
25	1\\
};

\addplot [color=mycolor1, thick]
  table[row sep=crcr]{%
1	0.0509575050608442\\
5	0.21518034394557\\
10	0.356945394777751\\
15	0.454121635775625\\
25	0.573664983311793\\
};
\addlegendentry{$\text{G = 8}\times\text{4}$}

\addplot [color=mycolor2, thick, densely dashed]
  table[row sep=crcr]{%
1	0.187756612719733\\
5	0.617998709062204\\
10	0.824339349383409\\
15	0.904813232171474\\
25	0.959141491554811\\
};
\addlegendentry{$\text{G = 8}\times\text{8}$}

\addplot [color=mycolor3, thick, dashdotted]
  table[row sep=crcr]{%
1	0.503583064935968\\
5	0.964577806483619\\
10	0.997873305812219\\
15	0.999731577845146\\
25	0.999969135544622\\
};
\addlegendentry{$\text{G = 64}\times\text{4}$}

\addplot [color=mycolor4, thick, dotted]
  table[row sep=crcr]{%
1	0.712019568837855\\
5	0.997540434638726\\
10	0.999988129423151\\
15	0.999999801915435\\
25	0.999999991468918\\
};
\addlegendentry{$\text{G = 256}\times\text{4}$}

\addplot [color=mycolor5, thick, dashed]
  table[row sep=crcr]{%
1	0.799519237852782\\
5	0.99952748719654\\
10	0.999999346463878\\
15	0.999999995717375\\
25	0.999999999865741\\
};
\addlegendentry{$\text{G = 256}\times\text{8}$}

\end{axis}

\begin{axis}[%
width=0.951\fwidth,
height=\fheight,
at={(0\fwidth,0\fheight)},
scale only axis,
xmin=0,
xmax=25,
xlabel style={font=\footnotesize\color{white!15!black}},
xlabel={Altitude $h$ [m]},
ymin=0,
ymax=1,
ylabel style={font=\footnotesize\color{white!15!black}},
ylabel={$P_{\rm cov}$},
xmajorgrids,
ymajorgrids,
legend style={legend cell align=left, align=left, draw=white!15!black, font=\footnotesize, at={(0.99, 0.01)}, anchor=south east},
hide y axis,
hide x axis,
]

\addplot [color=black]
  table[row sep=crcr]{%
-1 -1\\
};
\addlegendentry{Analysis}

\addplot [color=black, draw=none, mark size=1.3pt, mark=triangle*, mark options={solid, fill=black, black}, only marks]
  table[row sep=crcr]{%
-1 -1\\
};
\addlegendentry{Monte Carlo}
\end{axis}

\end{tikzpicture}%
    \setlength\belowcaptionskip{-.3cm}
    \caption{Coverage probability as a function of the \gls{uav} density $\lambda_{\rm UAV}$, for different configurations of the antenna arrays at the \glspl{uav} ($\mathcal{N}_{\rm UAV}$) and \glspl{ue} ($\mathcal{N}_{\rm UE}$) , with $\Gamma = 5$ dB and a fixed height $h=200$ m.}
    \label{fig:beamformingGain}
\end{figure}
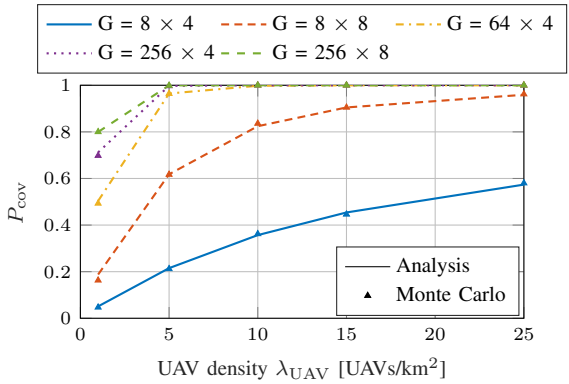

Fig.~\ref{fig:snr_thresh} reports the trend of $P_{\rm cov}$ for different values of the SNR threshold $\Gamma$, the UAV height $h$ and the UAV deployment density $\lambda_{\rm UAV}$.
The coverage probability exhibits a bell shape, with a peak at a specific value of the altitude $h_{\rm opt}$, which is given by the combination of two factors. On one hand, if the \glspl{uav} are deployed at low altitudes, the \gls{los} probability will be small, thus the increased pathloss in \gls{nlos} will penalize the probability of having an \gls{snr} above the threshold~$\Gamma$. Conversely, for high altitudes, even if the link is likely in \gls{los}, the impact of the increased distance between the reference UE and its serving \gls{uav} decreases the overall link budget, preventing a successful communication.

Notice that, as we do not model the interference in the coverage probability estimation, the performance of the network monotonically increases when considering larger values of the deployment density $\lambda_{\rm UAV}$. Notably, for $\Gamma = 0$ dB, only densities higher than 10 UAVs/km$^2$ reach $P_{\rm cov} \simeq 1$ at the optimal height. In general, the deployment with $\lambda_{\rm UAV} = 1$ UAVs/km$^2$ severely underperforms the others, with a gap of 0.12 (for $\Gamma=-5$ dB), 0.45 (for $\Gamma=0$ dB), and 0.55 (for $\Gamma=5$ dB) in the optimal coverage probability obtained with $\lambda_{\rm UAV}=5$ UAVs/km$^2$. With $\Gamma = -5$ or 0 dB, however, it does not make sense to increase the deployment density above 5 \gls{uav}/km$^2$ and 10 \gls{uav}/km$^2$, respectively, as $P_{\rm cov}$ saturates to 1 for the peak altitude.

Finally, when increasing $\Gamma$ (i.e., the minimum quality required to establish the link), the coverage probability decreases, as expected. In particular, while for $\Gamma = -5$ dB only the configuration with $\lambda_{\rm UAV}=1$ UAVs/km$^2$ does not reach $P_{\rm cov} = 1$, with $\Gamma = 5$ dB none of the configurations peaks at 1, with $\lambda_{\rm UAV} = 25$ UAVs/km$^2$ having a maximum value of 0.99. 
This is reasonable as an increase in the minimum acceptable SNR value for coverage translates into a tighter constraint on the link budget of the communication link. Additionally, it can be seen that, besides a decrease in the maximum value of $P_{\rm cov}$ for increasing values of $\Gamma$, the altitude for which the peak value is reached decreases, showing that the reduction in path gain has a more significant impact on Eq.~\eqref{Pcov_value_theorem} than the reduction of \gls{los} probability given by lower values of  $h$.

\subsection{Impact of the Antenna Configuration}

The configuration of the antenna array plays a fundamental role in reaching a high coverage probability while maintaining the highest SNR threshold $\Gamma = 5$ dB. The beamforming gain, indeed, is fundamental in mmWave communications to compensate for the larger pathloss that is experienced at such high frequencies. Fig.~\ref{fig:beamformingGain} shows the trend of $P_{\rm cov}$ when varying the \gls{uav} density $\lambda_{\rm UAV}$ and the number of antenna elements at the transmitter and the receiver. As can be seen, even for small values of $\lambda_{\rm UAV}$ (i.e., 5 UAVs/km$^2$), it is possible to achieve $P_{\rm cov} \ge 0.95$ by using at least 64 antenna elements in the \gls{uav} and only 4 in the \glspl{ue}. Similarly, by comparing the configurations with $\mathcal{N}_{\rm UAV} = 8$ (fixed) and with $\mathcal{N}_{\rm UE}$ varying from 4 to 8, it is possible to increase $P_{\rm cov}$ by 2 times for $\lambda_{\rm UAV}$ up to 10 UAVs/km$^2$.

\subsection{Impact of the Deployment Height}

\begin{figure}[t]
    \centering
    \setlength\fwidth{0.7\columnwidth}
    \setlength\fheight{0.35\columnwidth}
%
%
\definecolor{mycolor1}{rgb}{0.00000,0.44700,0.74100}%
\definecolor{mycolor2}{rgb}{0.85000,0.32500,0.09800}%
\definecolor{mycolor3}{rgb}{0.92900,0.69400,0.12500}%
\definecolor{mycolor4}{rgb}{0.49400,0.18400,0.55600}%
\definecolor{mycolor5}{rgb}{0.46600,0.67400,0.18800}%
\begin{tikzpicture}
\pgfplotsset{every tick label/.append style={font=\scriptsize}}

\begin{axis}[%
width=0.951\fwidth,
height=\fheight,
at={(0\fwidth,0\fheight)},
scale only axis,
xmin=0,
xmax=25,
xlabel style={font=\footnotesize\color{white!15!black}},
xlabel={UAV density $\lambda_{\rm UAV}$ [UAVs/km$^2$]},
ymin=0,
ymax=1,
ylabel style={font=\footnotesize\color{white!15!black}},
ylabel={$P_{\rm cov}$},
axis background/.style={fill=white},
xmajorgrids,
ymajorgrids,
legend style={at={(0.5,1.05)}, anchor=south, legend cell align=left, align=left, draw=white!15!black, font=\footnotesize},
legend columns=3,
]
\addplot [color=mycolor1, draw=none, mark size=1.3pt, mark=triangle*, mark options={solid, fill=mycolor1, mycolor1}, forget plot]
  table[row sep=crcr]{%
1	0.031\\
5	0.14\\
10	0.251\\
15	0.351\\
25	0.486\\
};

\addplot [color=mycolor2, draw=none, mark size=1.3pt, mark=triangle*, mark options={solid, fill=mycolor2, mycolor2}, forget plot]
  table[row sep=crcr]{%
1	0.106\\
5	0.446\\
10	0.724\\
15	0.848\\
25	0.955\\
};

\addplot [color=mycolor3, draw=none, mark size=1.3pt, mark=triangle*, mark options={solid, fill=mycolor3, mycolor3}, forget plot]
  table[row sep=crcr]{%
1	0.231\\
5	0.743\\
10	0.943\\
15	0.986\\
25	0.999\\
};

\addplot [color=mycolor4, draw=none, mark size=1.3pt, mark=triangle*, mark options={solid, fill=mycolor4, mycolor4}, forget plot]
  table[row sep=crcr]{%
1	0.36\\
5	0.894\\
10	0.989\\
15	0.999\\
25	1\\
};

\addplot [color=mycolor5, draw=none, mark size=1.3pt, mark=triangle*, mark options={solid, fill=mycolor5, mycolor5}, forget plot]
  table[row sep=crcr]{%
1	0.452\\
5	0.943\\
10	1\\
15	1\\
25	1\\
};

\addplot [color=mycolor1, thick]
  table[row sep=crcr]{%
1	0.0278167222842979\\
5	0.130979464008721\\
10	0.243553402529056\\
15	0.340467368681009\\
25	0.496211488507622\\
};
\addlegendentry{$h = 2$ m}

\addplot [color=mycolor2, thick, densely dashed]
  table[row sep=crcr]{%
1	0.114920976620632\\
5	0.45556376419224\\
10	0.701813421619297\\
15	0.835704141970833\\
25	0.949218070247852\\
};
\addlegendentry{$h = 52$ m}

\addplot [color=mycolor3, thick, dashdotted]
  table[row sep=crcr]{%
1	0.24967780262429\\
5	0.759571804662879\\
10	0.94052611716028\\
15	0.984830566420579\\
25	0.998905343721597\\
};
\addlegendentry{$h = 102$ m}

\addplot [color=mycolor4, thick, dotted]
  table[row sep=crcr]{%
1	0.372991214294411\\
5	0.89841732955693\\
10	0.98816890606077\\
15	0.998348452990236\\
25	0.999927965307999\\
};
\addlegendentry{$h = 152$ m}

\addplot [color=mycolor5, thick, dashed]
  table[row sep=crcr]{%
1	0.458111243791103\\
5	0.945149520418434\\
10	0.995045420259482\\
15	0.999159891290942\\
25	0.999882026216246\\
};
\addlegendentry{$h = 202$ m}

\end{axis}

\begin{axis}[%
width=0.951\fwidth,
height=\fheight,
at={(0\fwidth,0\fheight)},
scale only axis,
xmin=0,
xmax=25,
xlabel style={font=\footnotesize\color{white!15!black}},
xlabel={Altitude $h$ [m]},
ymin=0,
ymax=1,
ylabel style={font=\footnotesize\color{white!15!black}},
ylabel={$P_{\rm cov}$},
xmajorgrids,
ymajorgrids,
legend style={legend cell align=left, align=left, draw=white!15!black, font=\footnotesize, at={(0.99, 0.01)}, anchor=south east},
hide y axis,
hide x axis,
]

\addplot [color=black]
  table[row sep=crcr]{%
-1 -1\\
};
\addlegendentry{Analysis}

\addplot [color=black, draw=none, mark size=1.3pt, mark=triangle*, mark options={solid, fill=black, black}, only marks]
  table[row sep=crcr]{%
-1 -1\\
};
\addlegendentry{Monte Carlo}
\end{axis}
\end{tikzpicture}%
    \setlength\belowcaptionskip{-.3cm}
    \caption{Coverage probability as a function of the \gls{uav} density $\lambda_{\rm UAV}$, for different values of the altitude of the deployment, $\mathcal{N}_{\rm UAV}=8$ and $\mathcal{N}_{\rm UE}=8$, and $\Gamma = 0$ dB.}
    \label{fig:height}
\end{figure}
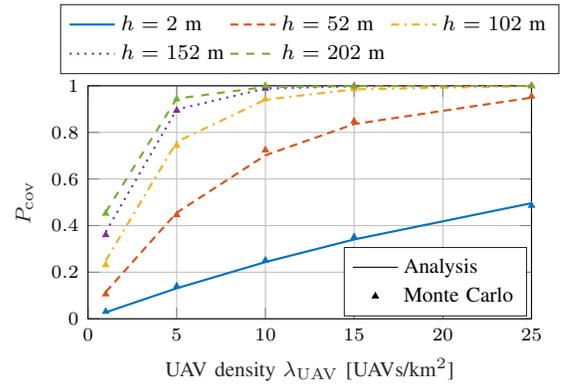

As highlighted in Fig.~\ref{fig:snr_thresh}, the height at which the \glspl{uav} are deployed is a key configuration parameter of the network. This parameter distinguishes \gls{uav} networks from traditional cellular networks, as the latter have constraints on the base station height (e.g., the availability of poles and towers) that are more relaxed when operating \glspl{uav}. Fig.~\ref{fig:height} reports the values of $P_{\rm cov}$ for different values of $\lambda_{\rm UAV}$ and deployment height $h$. The latter varies between 2 and 202 m, i.e., in a range where the maximum value is smaller than the height at which $P_{\rm cov}$ peaks and then decreases.\footnote{Notice that several civil flight authorities limit the height of \glspl{uav} to values comparable to those selected in Fig.~\ref{fig:height}~\cite{galkin2019stochastic}.} Fig.~\ref{fig:height} shows that, by operating at a higher altitude, it is possible to decrease the deployment density of the \glspl{uav}, without compromising the coverage probability. For example, the configurations with $\lambda_{\rm UAV}=1$ UAVs/km$^2$, and $h=202$~m, has a similar $P_{\rm cov}$ to that with $h=2$ m and $\lambda_{\rm UAV}=10$ UAVs/km$^2$. Notice that, in terms of deployment cost and energy, it is more efficient to operate fewer \glspl{uav}, even if placed at a higher altitude. Along this line, given the range of values of $\lambda_{\rm UAV}$ considered in Fig.~\ref{fig:height}, it is possible to reach the maximum value for $P_{\rm cov}$ only with the highest-altitude deployments.

\section{Conclusions and Future Work} 
\label{sec:conclusions_and_future_work}
In this work, we proposed a stochastic geometry framework to characterize the coverage  performance of a mmWave UAV network.
Our model characterizes the UAVs as two independent LOS and NLOS two-dimensional homogeneous Poisson processes and derives the analytical expression for the SNR coverage probability as a function of the distance between a reference  ground user terminal and its serving UAV. 
We validated our analytical curves with Monte Carlo simulations and  investigated the relation between coverage support and UAV altitude, UAV density, antenna architecture and targeted link quality threshold.
We demonstrated that there exists an optimal altitude at which the UAV should be placed to satisfy the required signal quality threshold, which depends on the UAV swarm size and beamforming gain.

UAV networks still suffer from several challenges including energy consumption, reduced time of flight and environment-related sensitivity. In particular, future research efforts should  be dedicated to discovering the relationship between coverage performance and UAV energy consumption.

\appendices
\section{}
\label{appendix:f_i}
The distance from the closest UAV of type $i\in\{L,N\}$ is equal to $r$ if there are no other UAVs of the same type closer than $r$. Taking as a reference the model in Fig.~\ref{fig:model}, this means that there must be no other  UAV of type  $i$ in the ball $\mathcal{B}(0,b(r))$, where we can define the  positions of the UAVs in $\mathcal{B}(0,b(r))$ in polar coordinates as a function of the variables $\rho$ and $\xi$, with $0 \leq \rho \leq b(r)$ and $0 \leq \xi \leq 2\pi$. Now, since the process $\Phi_{\text{UAV},i}$ is a 2D homogeneous PPP with intensity measure $\lambda_{\text{UAV},i}$ over $\mathcal{B}(0,b(r))$, we get
\begin{align}
\label{eq:F_i}
F_i(r)&=\mathbb{P}\Big[\text{No LOS UAVs in the ball } \mathcal{B}(0,b(r))\Big] \notag  \\ 
&=\exp\left(-\int_0^{2\pi} \int_0^{b(r)} \lambda_{\text{UAV},i} \rho d\rho d\xi\right)\notag\\ 
&\stackrel{(a)}= \exp\left(- 2\pi \lambda_{\rm UAV} \int_0^{b(r)} p_i(\rho)\rho d\rho \right)
\end{align}
with $b(r) = \sqrt{r^2-h^2}$ and $(a)$ derives from $\lambda_{\text{UAV},i} =p_i(\rho)\lambda_{\rm UAV}$. In Eq.~\eqref{eq:F_i}, $p_i(\rho)$ can be rewritten from~\eqref{plosoverrho} as a function of $\rho$ as
\begin{align}\label{plosoverr}
    p_{i}(\rho) = \frac{1}{1+C\exp\left[-Y\left(\arctan\left(\frac{h}{\rho}\right)\frac{180}{\pi}-C\right)\right]}.
\end{align}
 The \gls{pdf} of the distance $r$ to the closest UAV of type $i$ can finally be computed as
\begin{align}
&f_i(r) = \frac{\partial }{ \partial r} \Big(1- F_i(r)\Big) \notag \\
&= 2\pi\lambda_{\rm UAV}rp_i(b(r))\exp\left(-2\pi\lambda_{\rm UAV}\int_0^{b(r)}p_i(\rho)\rho d\rho\right).
\end{align}

\section{}
\label{appendix:P_i}

Let $r_L$ and $r_N$ be random variables expressing the distance of the reference ground UE from the closest LOS and NLOS UAVs, respectively. In order for the user to connect with a LOS (NLOS) UAV, there must be no NLOS (LOS) UAVs at a distance smaller than or equal to $A_L(r)$ ($A_N(r)$), i.e., the minimum distance at which a potential NLOS (LOS) UAV could provide a better path gain to the user in comparison to its LOS rival.
Considering the LOS case, the probability $P_L$ of connecting to a LOS UAV can be written as
\begin{align}
\label{eq:P_i_1}
        P_L(r) &= \mathbb{P}\left[C_Lr_L^{-a_L} > C_N r_N^{-a_N}\right] \notag \\
        &= \mathbb{P}\left[r_N > \left(({C_N}/{C_L})r^{a_L}\right)^{\frac{1}{a_N}}\right] \notag \\
        &\stackrel{(a)}= \int_h^\infty \mathbb{P}\left[r_N>A_L(r)\right]f_L(r)dr
\end{align}
where $(a)$ follows from the fact that $r>h$ by construction,  and $f_L$ is the \gls{pdf} of $r_L$ according to Lemma~\ref{f_i}.
In Eq.~\eqref{eq:P_i_1}, $\mathbb{P}\left[r_N>A_L(r)\right]$ represents the probability that there are no NLOS BS inside the circular area $\mathcal{B}(0,b(A_L(r)))$ with radius $b(A_L(r))$ and can be computed as
\begin{align}
\label{eq:P_i_2}
\mathbb{P}\left[r_N>A_L(r)\right] = \exp\left(-2\pi \lambda_{\rm UAV} \int_0^{b\left(A_L(r)\right)} p_{N}(\rho) \rho d\rho\right),
\end{align}
where $p_{N}(\rho)$ follows from Eq.~\eqref{plosoverr}. By substituting~\eqref{eq:P_i_2} into~\eqref{eq:P_i_1}, we get the expression in Lemma~\ref{P_i} for the LOS case. With a similar proof, it is possible to prove the lemma also for the NLOS case. 
        
\section{}
\label{appendix:P_cov}
Being $r_i$ the distance between the reference ground UE and its serving UAV $n^*\in\Phi_{\text{UAV},i}$, with $i\in\{L,N\}$, the probability in Eq.~\eqref{eq:SNR_divided} can be written as
\begin{align}\label{P_cov_1}
            \mathbb{P}&\Big[\text{SNR}_i > \Gamma, n^* \in \Phi_i\Big] = \mathbb{E}_{r_i}\Bigg[\mathbb{P}\Big[\text{SNR}_i > \Gamma \Bigl\lvert r_i\Big]\Bigg] \notag \\
            &\stackrel{(a)}=\mathbb{E}_{r_i}\Bigg[\mathbb{P}\left[\frac{P_{\rm TX}\ell_i(r_i)Gg(m_i,\Omega_i)}{\text{NF}\cdot\sigma^2} > \Gamma \Bigl\lvert r_i\right]\Bigg] \notag \\
            &= \int_h^{\infty} \mathbb{P}\left[g(m_i,\Omega_i) > \frac{\Gamma(\text{NF}\cdot\sigma^2)}{P_{\rm TX}GC_i}r^{a_i} \Bigl\lvert r\right]\overline{f}_i(r)dr,
\end{align}
where $(a)$ derives from Eq.~\eqref{eq:SNR}. 
Now, with the assumption that the small scale fading $g$ follows a Nakagami distribution of parameters $m_i$ and $\Omega_i$, $i\in\{L,N\}$, we can derive  its associated \gls{cdf} as
\begin{align}
    F(x;m_i,\Omega_i) &= \mathbb{P}\Big[g(m_i,\Omega_i) \leq x\Big] \notag \\
    &= P\left(m_i,\frac{m_i}{\Omega_i}x^2\right) = \frac{\gamma(m_i,\frac{m_i}{\Omega_i}x^2)}{\Gamma(m_i)},
\end{align}
where $P\left(m,\frac{m_i}{\Omega_i}x^2\right)$ is the (regularized) incomplete gamma function, $\gamma(m_i,\frac{m_i}{\Omega_i}x^2)$ is the lower incomplete gamma function and $\Gamma(m_i) = \left(m_i-1\right)!$.
Therefore, the probability term inside Eq.~\eqref{P_cov_1} becomes
\begin{align}\label{P_cov_2}
 &\mathbb{P}\left[g(m_i,\Omega_i) > \frac{\Gamma(\text{NF}\cdot\sigma^2)}{P_{\rm TX}GC_i}r^{a_i} \Bigl\lvert r\right] \notag \\
 &=1-P\left(m_i,\frac{m_i}{\Omega_i}\zeta_i(r)^2\right) = 1-\frac{\int_0^{\Delta_i}t^{m_i-1}e^{-t}dt}{\left(m_i-1\right)!},
\end{align}
where we defined $\zeta_i(r) = \frac{\Gamma(\text{NF}\cdot\sigma^2)}{P_{\rm TX}GC_i}r^{a_i}$ and $\Delta_i = \frac{m_i}{\Omega_i}\zeta_i(r)^2$.\\ 
By substituting \eqref{P_cov_2} into \eqref{P_cov_1} we get the expression of the coverage probability in Theorem~\ref{th:pcov}.

\bibliographystyle{IEEEtran}
\bibliography{bibliography.bib}

\begin{thebibliography}{10}
\providecommand{\url}[1]{#1}
\csname url@samestyle\endcsname
\providecommand{\newblock}{\relax}
\providecommand{\bibinfo}[2]{#2}
\providecommand{\BIBentrySTDinterwordspacing}{\spaceskip=0pt\relax}
\providecommand{\BIBentryALTinterwordstretchfactor}{4}
\providecommand{\BIBentryALTinterwordspacing}{\spaceskip=\fontdimen2\font plus
\BIBentryALTinterwordstretchfactor\fontdimen3\font minus
  \fontdimen4\font\relax}
\providecommand{\BIBforeignlanguage}[2]{{%
\expandafter\ifx\csname l@#1\endcsname\relax
\typeout{** WARNING: IEEEtran.bst: No hyphenation pattern has been}%
\typeout{** loaded for the language `#1'. Using the pattern for}%
\typeout{** the default language instead.}%
\else
\language=\csname l@#1\endcsname
\fi
#2}}
\providecommand{\BIBdecl}{\relax}
\BIBdecl

\bibitem{chandhar2017massive}
P.~Chandhar, D.~Danev, and E.~G. Larsson, ``{Massive MIMO for communications
  with drone swarms},'' \emph{IEEE Transactions on Wireless Communications},
  vol.~17, no.~3, pp. 1604--1629, Mar 2017.

\bibitem{chandhar2019massive}
P.~Chandhar and E.~G. Larsson, ``{Massive MIMO for connectivity with drones:
  Case studies and future directions},'' \emph{IEEE Access}, vol.~7, pp.
  94\,676--94\,691, 2019.

\bibitem{shi2018drone}
W.~Shi, H.~Zhou, J.~Li, W.~Xu, N.~Zhang, and X.~Shen, ``{Drone assisted
  vehicular networks: Architecture, challenges and opportunities},'' \emph{IEEE
  Network}, vol.~32, no.~3, pp. 130--137, May/June 2018.

\bibitem{mezzavilla2018public}
M.~{Mezzavilla}, M.~{Polese}, A.~{Zanella}, A.~{Dhananjay}, S.~{Rangan},
  C.~{Kessler}, T.~S. {Rappaport}, and M.~{Zorzi}, ``{Public Safety
  Communications above 6 GHz: Challenges and Opportunities},'' \emph{IEEE
  Access}, vol.~6, pp. 316--329, 2018.

\bibitem{giordani2020satellite}
M.~Giordani and M.~Zorzi, ``{Satellite Communication at Millimeter Waves: a Key
  Enabler of the 6G Era},'' \emph{IEEE International Conference on Computing,
  Networking and Communications (ICNC)}, Feb 2020.

\bibitem{van2016lte}
B.~Van Der~Bergh, A.~Chiumento, and S.~Pollin, ``{LTE in the sky: Trading off
  propagation benefits with interference costs for aerial nodes},'' \emph{IEEE
  Communications Magazine}, vol.~54, no.~5, pp. 44--50, May 2016.

\bibitem{hayat2015experimental}
S.~Hayat, E.~Yanmaz, and C.~Bettstetter, ``{Experimental analysis of
  multipoint-to-point UAV communications with IEEE 802.11n and 802.11ac},'' in
  \emph{IEEE 26th Annual International Symposium on Personal, Indoor, and
  Mobile Radio Communications (PIMRC)}.\hskip 1em plus 0.5em minus 0.4em\relax
  IEEE, 2015, pp. 1991--1996.

\bibitem{zhang2019survey}
L.~Zhang, H.~Zhao, S.~Hou, Z.~Zhao, H.~Xu, X.~Wu, Q.~Wu, and R.~Zhang, ``{A
  survey on 5G millimeter wave communications for UAV-assisted wireless
  networks},'' \emph{IEEE Access}, vol.~7, pp. 117\,460--117\,504, 2019.

\bibitem{xiao2017millimeter}
M.~Xiao, S.~Mumtaz, Y.~Huang, L.~Dai, Y.~Li, M.~Matthaiou, G.~K. Karagiannidis,
  E.~Bj{\"o}rnson, K.~Yang, I.~Chih-Lin \emph{et~al.}, ``{Millimeter wave
  communications for future mobile networks},'' \emph{IEEE Journal on Selected
  Areas in Communications}, vol.~35, no.~9, pp. 1909--1935, Sept 2017.

\bibitem{rappaport2014millimeter}
T.~S. Rappaport, R.~W. Heath~Jr, R.~C. Daniels, and J.~N. Murdock,
  \emph{Millimeter wave wireless communications}.\hskip 1em plus 0.5em minus
  0.4em\relax Pearson Education, 2014.

\bibitem{giordani2018tutorial}
M.~Giordani, M.~Polese, A.~Roy, D.~Castor, and M.~Zorzi, ``{A tutorial on beam
  management for 3GPP NR at mmWave frequencies},'' \emph{IEEE Communications
  Surveys \& Tutorials}, vol.~21, no.~1, pp. 173--196, First quarter 2019.

\bibitem{liu2018energy}
C.~H. Liu, Z.~Chen, J.~Tang, J.~Xu, and C.~Piao, ``{Energy-efficient UAV
  control for effective and fair communication coverage: A deep reinforcement
  learning approach},'' \emph{IEEE Journal on Selected Areas in
  Communications}, vol.~36, no.~9, pp. 2059--2070, Sept 2018.

\bibitem{zhang2018cellular}
S.~Zhang, Y.~Zeng, and R.~Zhang, ``{Cellular-enabled UAV communication: A
  connectivity-constrained trajectory optimization perspective},'' \emph{IEEE
  Transactions on Communications}, vol.~67, no.~3, pp. 2580--2604, Mar 2019.

\bibitem{36777}
3GPP, ``{Study on enhanced LTE support for aerial vehicles},'' TR 36.777,
  V15.0.0, 2017.

\bibitem{22125}
------, ``{Unmanned Aerial System (UAS) support in 3GPP},'' TR 22.125, V17.1.0,
  2019.

\bibitem{al2014optimal}
A.~Al-Hourani, S.~Kandeepan, and S.~Lardner, ``{Optimal LAP altitude for
  Maximum Coverage},'' \emph{IEEE Wireless Communications Letters}, vol.~3,
  no.~6, pp. 569--572, Dec 2014.

\bibitem{ahmed2016importance}
N.~Ahmed, S.~S. Kanhere, and S.~Jha, ``{On the importance of link
  characterization for aerial wireless sensor networks},'' \emph{IEEE
  Communications Magazine}, vol.~54, no.~5, pp. 52--57, May 2016.

\bibitem{mozaffari2015drone}
M.~Mozaffari, W.~Saad, M.~Bennis, and M.~Debbah, ``{Drone small cells in the
  clouds: Design, deployment and performance analysis},'' in \emph{IEEE Global
  Communications Conference (GLOBECOM)}.\hskip 1em plus 0.5em minus 0.4em\relax
  IEEE, 2015.

\bibitem{bertizzolo2019mmbac}
L.~Bertizzolo, M.~Polese, L.~Bonati, A.~Gosain, M.~Zorzi, and T.~Melodia,
  ``{mmBAC: Location-Aided mmWave Backhaul Management for UAV-Based Aerial
  Cells},'' in \emph{3rd ACM Workshop on Millimeter-Wave Networks and Sensing
  Systems}, ser. mmNets’19.\hskip 1em plus 0.5em minus 0.4em\relax
  Association for Computing Machinery, 2019.

\bibitem{cuverlier2018mmWaveAerial}
T.~Cuvelier and R.~Heath, ``{mmWave} {MU-MIMO} for aerial networks,'' in
  \emph{Proc. of IEEE ISWCS}, Lisbon, Portugal, Aug 2018.

\bibitem{xia2019millimeter}
W.~{Xia}, M.~{Polese}, M.~{Mezzavilla}, G.~{Loianno}, S.~{Rangan}, and
  M.~{Zorzi}, ``{Millimeter Wave Remote UAV Control and Communications for
  Public Safety Scenarios},'' in \emph{16th Annual IEEE International
  Conference on Sensing, Communication, and Networking (SECON)}, June 2019.

\bibitem{xiao2016enabling}
Z.~Xiao, P.~Xia, and X.~Xia, ``Enabling {UAV} cellular with millimeter-wave
  communication: Potentials and approaches,'' \emph{IEEE Communications
  Magazine}, vol.~54, no.~5, pp. 66--73, May 2016.

\bibitem{gapeyenko2018flexible}
M.~{Gapeyenko}, V.~{Petrov}, D.~{Moltchanov}, S.~{Andreev}, N.~{Himayat}, and
  Y.~{Koucheryavy}, ``{Flexible and Reliable UAV-Assisted Backhaul Operation in
  5G mmWave Cellular Networks},'' \emph{IEEE Journal on Selected Areas in
  Communications}, vol.~36, no.~11, pp. 2486--2496, Nov 2018.

\bibitem{khawaja2017mmWaveUAVChannel}
W.~Khawaja, O.~Ozdemir, and I.~Guvenc, ``{UAV} air-to-ground channel
  characterization for {mmWave} systems,'' in \emph{Proc. of IEEE VTC-Fall},
  Toronto, ON, Canada, Sept 2017.

\bibitem{khawaja2018temporal}
------, ``Temporal and spatial characteristics of {mmWave} propagation channels
  for {UAVs},'' in \emph{Proc. of IEEE GSMM}, May 2018.

\bibitem{zhao2018channel}
J.~Zhao, G.~Gao, L.~Kuang, Q.~Wu, and W.~Jia, ``Channel tracking with flight
  control system for {UAV} {mmWave} {MIMO} communications,'' \emph{IEEE
  Communications Letters}, vol.~22, no.~6, pp. 1224--1227, June 2018.

\bibitem{ravi2016downlink}
V.~V.~C. Ravi and H.~S. Dhillon, ``{Downlink coverage probability in a finite
  network of unmanned aerial vehicle (UAV) base stations},'' in \emph{IEEE 17th
  International Workshop on Signal Processing Advances in Wireless
  Communications (SPAWC)}.\hskip 1em plus 0.5em minus 0.4em\relax IEEE, 2016,
  pp. 1--5.

\bibitem{chetlur2017downlink}
V.~V. Chetlur and H.~S. Dhillon, ``{Downlink coverage analysis for a finite 3-D
  wireless network of unmanned aerial vehicles},'' \emph{IEEE Transactions on
  Communications}, vol.~65, no.~10, pp. 4543--4558, Oct 2017.

\bibitem{galkin2017stochastic}
B.~Galkin, J.~Kibi{\l}da, and L.~A. DaSilva, ``{A stochastic geometry model of
  backhaul and user coverage in urban UAV networks},'' \emph{arXiv preprint
  arXiv:1710.03701}, 2017.

\bibitem{zhang2016spectrum}
C.~Zhang and W.~Zhang, ``{Spectrum sharing for drone networks},'' \emph{IEEE
  Journal on Selected Areas in Communications}, vol.~35, no.~1, pp. 136--144,
  Jan 2017.

\bibitem{azari2017coverage}
M.~M. Azari, Y.~Murillo, O.~Amin, F.~Rosas, M.-S. Alouini, and S.~Pollin,
  ``{Coverage maximization for a poisson field of drone cells},'' in \emph{IEEE
  28th Annual International Symposium on Personal, Indoor, and Mobile Radio
  Communications (PIMRC)}.\hskip 1em plus 0.5em minus 0.4em\relax IEEE, 2017.

\bibitem{liu2018performance}
C.~Liu, M.~Ding, C.~Ma, Q.~Li, Z.~Lin, and Y.-C. Liang, ``{Performance analysis
  for practical unmanned aerial vehicle networks with LoS/NLoS
  transmissions},'' in \emph{IEEE International Conference on Communications
  Workshops (ICC Workshops)}.\hskip 1em plus 0.5em minus 0.4em\relax IEEE,
  2018, pp. 1--6.

\bibitem{galkin2018backhaul}
B.~Galkin, J.~Kibilda, and L.~A. DaSilva, ``{Backhaul for low-altitude UAVs in
  urban environments},'' in \emph{IEEE International Conference on
  Communications (ICC)}.\hskip 1em plus 0.5em minus 0.4em\relax IEEE, 2018.

\bibitem{hayajneh2016drone}
A.~M. Hayajneh, S.~A.~R. Zaidi, D.~C. McLernon, and M.~Ghogho, ``{Drone
  empowered small cellular disaster recovery networks for resilient smart
  cities},'' in \emph{IEEE international conference on sensing, communication
  and networking (SECON Workshops)}.\hskip 1em plus 0.5em minus 0.4em\relax
  IEEE, 2016.

\bibitem{galkin2019stochastic}
B.~Galkin, J.~Kibi{\l}da, and L.~A. DaSilva, ``{A stochastic model for UAV
  networks positioned above demand hotspots in urban environments},''
  \emph{IEEE Transactions on Vehicular Technology}, vol.~68, no.~7, pp.
  6985--6996, July 2019.

\bibitem{zhou2018underlay}
X.~Zhou, S.~Durrani, J.~Guo, and H.~Yanikomeroglu, ``{Underlay drone cell for
  temporary events: Impact of drone height and aerial channel environments},''
  \emph{IEEE Internet of Things Journal}, vol.~6, no.~2, pp. 1704--1718, Apr
  2019.

\bibitem{Bacelli_book}
F.~Baccelli and B.~Błaszczyszyn, ``{Stochastic Geometry and Wireless Networks:
  Volume {I} Theory},'' \emph{{Foundations and Trends® in Networking}},
  vol.~3, no. 3–4, pp. 249--449, 2010.

\bibitem{Coverage_rate_analysis}
T.~Bai and R.~W. Heath, ``{Coverage and Rate Analysis for Millimeter-Wave
  Cellular Networks},'' \emph{IEEE Transactions on Wireless Communications},
  vol.~14, no.~2, pp. 1100--1114, Feb 2015.

\bibitem{giordani2018coverage}
M.~Giordani, M.~Rebato, A.~Zanella, and M.~Zorzi, ``Coverage and connectivity
  analysis of millimeter wave vehicular networks,'' \emph{Ad Hoc Networks},
  vol.~80, pp. 158--171, Aug 2018.

\bibitem{zhu2018secrecy}
Y.~{Zhu}, G.~{Zheng}, and M.~{Fitch}, ``{Secrecy Rate Analysis of UAV-Enabled
  mmWave Networks Using Matérn Hardcore Point Processes},'' \emph{IEEE Journal
  on Selected Areas in Communications}, vol.~36, no.~7, pp. 1397--1409, July
  2018.

\bibitem{andrews2017modeling}
J.~G. {Andrews}, T.~{Bai}, M.~N. {Kulkarni}, A.~{Alkhateeb}, A.~K. {Gupta}, and
  R.~W. {Heath}, ``Modeling and analyzing millimeter wave cellular systems,''
  \emph{IEEE Transactions on Communications}, vol.~65, no.~1, pp. 403--430, Jan
  2017.

\bibitem{Mustafa}
M.~R. Akdeniz, Y.~Liu, M.~K. Samimi, S.~Sun, S.~Rangan, T.~S. Rappaport, and
  E.~Erkip, ``{Millimeter Wave Channel Modeling and Cellular Capacity
  Evaluation},'' \emph{IEEE Journal on Selected Areas in Communications},
  vol.~32, no.~6, pp. 1164--1179, June 2014.

\end{thebibliography}

\end{document}